\documentclass[a4paper,11pt]{article}
\usepackage[utf8]{inputenc}
\usepackage[T1]{fontenc}
\usepackage[english]{babel}

\usepackage{graphicx}
\graphicspath{{Figures/}}
\usepackage{amsmath}
\usepackage{amssymb}
\usepackage{latexsym}
\usepackage[colorlinks]{hyperref}
\usepackage{color}
\usepackage{float}
\usepackage{cite}
\usepackage{makeidx}
\usepackage{colortbl}
\usepackage{csquotes}

\newcommand{\bra}{\begin{array}}
\newcommand{\era}{\end{array}}
\newcommand{\beq}{\begin{equation}}
\newcommand{\eeq}{\end{equation}}
\newcommand{\beqar}{\begin{eqnarray}}
\newcommand{\eeqar}{\end{eqnarray}}

\def\BC{\bb C}
\def\_\BC{\bbi C}

\def\( {\left(}
   \def\) {\right)}
\def\[ {\left[}
\def\] {\right]}
\def\no2 {{\textstyle{n\over 2}}}

\newcommand{\si}{\sigma}

\newcommand{\te}{\theta}

\newcommand{\lb}{\label}

\begin{document}
\begin{titlepage}
\setcounter{page}{1}
\renewcommand{\thefootnote}{\fnsymbol{footnote}}
\begin{flushright}
\end{flushright}
\vspace{5mm}
\begin{center}
{\Large \bf
Confinement in Gapped Graphene  with Magnetic Flux 
}

\vspace{5mm}

{\bf Bouchaib Lemaalem}$^{a}$, {\bf Abdelhadi Belouad}$^{a}$, {\bf Miloud Mekkaoui}$^{a}$ and  {\bf Ahmed Jellal\footnote{\sf 
a.jellal@ucd.ac.ma}}$^{a,b}$
\vspace{5mm}

{$^{a}$\em Laboratory of Theoretical Physics,
Faculty of Sciences, Choua\"ib Doukkali University},\\
{\em PO Box 20, 24000 El Jadida, Morocco}

{$^{b}$\em Canadian Quantum  Research Center,
204-3002 32 Ave Vernon, \\ BC V1T 2L7,  Canada}

\vspace{3cm}
\begin{abstract}
We study the propagation of electrons in a circular quantum dot
of gapped graphene subject to the magnetic flux  $\phi$. We present analytical expressions for the eigenstates, scattering  coefficients, scattering  efficiency and  radial component of the  reflected current. We identify different scattering regimes as a function of the physical parameters such as the incident electronic energy, potential barrier, radius of quantum dot,  gap and $\phi$.
We choose two values of the flux $\phi=1/2, 3/2$
and show that for  low energy of the incident electron, the scattering resonances appear and the far-field scattered current presents distinct preferred scattering directions.

\end{abstract}

\end{center}
\vspace{5cm}



\end{titlepage}

\section{Introduction}

Graphene is a mono-atomic layer of carbon atoms arranged on a two-dimensional honeycomb network and embedded in hexagonal structure
\cite{K. S. Novoselov2004,{K. S. Novoselov2005}}.
 Since its first isolation, graphene has rapidly changed its status from being unexpected and sometimes unwanted newcomer \cite{A. K. Geim12009} to a potential candidate to replace the semiconductors in electronic devices.
Because of
the exostic  properties quite different
to other materials in condensed matter physics, graphene
 has been attracting unfading interest as a fascinating system for fundamental studies.  
The optical and electronic transport properties of graphene nanostructures
\cite{S.Stander2009,S.G.Nam2011}
have been widely studied in recent years because they allow probing the fundamental laws of quantum physics. Such properties could also find applications in nanoelectronics as detectors and transistors \cite{A.H.Castro Neto2009} as well as present 
graphene as promising and  perspective material for different technological usages in future
\cite{A. K. Geim2007}.

Graphene quantum dots (QDs) 
exhibit extraordinary properties due to the quantum
confinement and edge effects \cite{L.A.Ponomarenkol2008}
because they
 have  useful optical and electronic properties associated with their nanometric structures.
QDs have been identified as attractive candidates for spin qubits and the storage of quantum information \cite{M.I.Katsnelson2006,J.Cserti2007}
as well as
in the fields of bio-imaging, sensors, catalysis,
photovoltaic devices, supercapacitors, etc \cite{333}.
The confinement of  Dirac fermions via an electrostatic grid potential is a difficult task because
the Klein effect allows massless fermions to tunnel through any electrostatic potential barrier \cite{M.M.Asmar2013}.
Such situation can be raised by manipulating the carrier states in graphene either using
an infinite graphene sheet or applying an external magnetic field
\cite{444}.
Graphene QDs can be realized by cutting small flakes from a graphene sheet \cite{J.S.Wu2014} or using the substrate to induce a
band gap \cite{2323}. QDs have energy states
that are strongly dependent on the shape, size and type of the boundary edges \cite{J.H.Bardarson2009,
G.Pa201l,R.Heinisch2013,A.Pieper2013,J. Wurm2009,P.Recher}.
It was argued    that such states 
with respect to QD size and external magnetic field are quantitatively similar
to each other \cite{888}.
On the other hand, the influence of a magnetic flux tube on the possibility to electrostatically confine fermions in a
graphene quantum dot was studied in \cite{Heinl2013}. Indeed, it was showed that without a magnetic flux tube, the graphene pseudospin is responsible for a quantization
of the total angular momentum to half-integer values. However,
with a flux tube containing half
a flux quantum
it was found that a state at zero
angular momentum that cannot be confined electrostatically.

Beside graphene sheet and confining potential together with energy gap,
we suggest
that magnetic flux $\phi$ could also be used to manipulate the carrier state
energies in this system.
%
Then, based on the findings in \cite{Heinl2013} and our previous work
\cite{Belouad2018},  we study the propagation of electrons in a circular quantum dot defined electrostatically in graphene according to an energy gap inside the quantum dot in the presence of 
the flux $\phi$. We identify different scattering regimes depending on the radius, potential, energy gap, incident energy and $\phi$. Using the boundary condition, we determine the scattering coefficients and the far-field radial component of the  reflected current associated to the reflected wave. 
For the two values $\phi=1/2, 3/2$,
we present numerical study by showing that  $\phi$ affects  the  scattering efficiency
and reflected current compared to  the case of without flux  \cite{Belouad2018}. In particular, under suitable conditions, these quantities present different oscillatory behaviors and resonances together with sharp picks at resonance points.

The paper is organized as follows. In section 2, we set a  theoretical model that allows us to describe the wave plane propagation 
in a circular quantum dot of  gaped graphene subject to a magnetic flux.
The solutions of the energy spectrum resulted from Dirac equation
are determined by considering two regions and the scattering coefficients are obtained via the boundary
condition.
In section {3}, we calculate the scattering efficiency
and the far-field radial component of the reflected current in terms
of the magnetic flux.
Section {4} will be devoted to  the  discussions  of our numerical results
and comparisons with the case $\phi=0$.
We conclude our work in the final section.

\section{Band structures}

\begin{figure}[!hbt]
\centering
\includegraphics[scale=0.25]{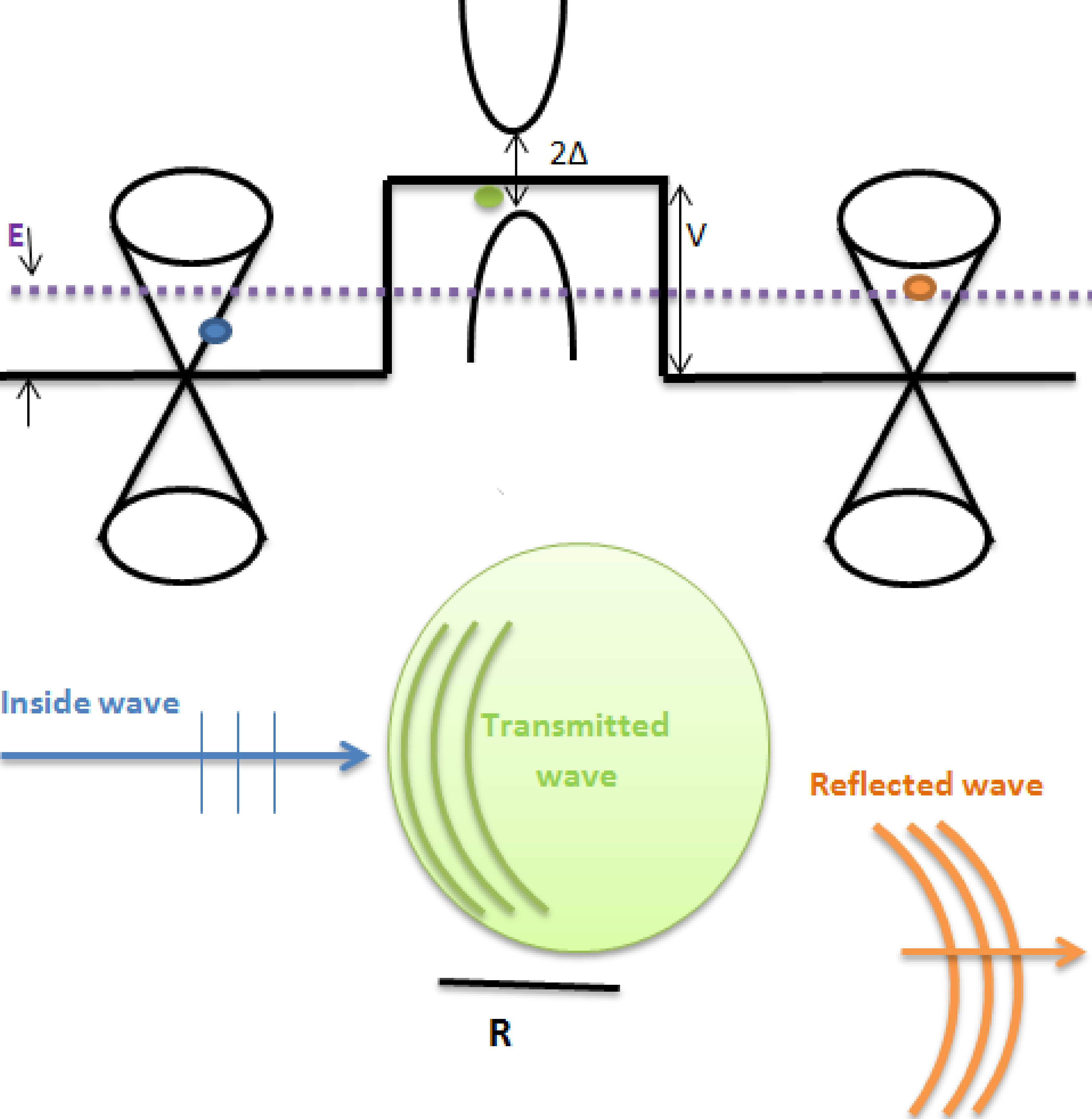}\\
\caption{\sf (color online) 
Dirac electrons propagating in a graphene  sheet subject to  the potential barrier  in a circular quantum dot of radius $R$,  gap $\Delta$ and magnetic flux $\phi$. The incident  and reflected electron waves reside in the conduction band, while the transmitted  wave  inside the dot corresponds to a state in the valence band.}\label{f0}
\end{figure}

To do our job, we formulate our theoretical model 
describing
the effect of the magnetic flux on
Dirac electrons in a circular electrostatically defined quantum dot of radius $R$ in
gapped graphene. Because 
 in graphene,
fermions are assigned a pseudospin associated  to the
sublattice degree of freedom,  then we consider 
 the single valley Hamiltonian 
\begin{equation}\label{e1}
 H= -i v_F \vec \sigma \cdot (\vec \nabla + e\vec A)   + \Delta(r)\sigma_z + V(r)\mathbb{I}
\end{equation}
involving the 
 potential barrier $V(r)$, energy gap $\Delta(r)$ and 
 vector potential $\vec A(r)$ associated to 
the magnetic flux   $\phi$ measured in  quantum unit
$\frac{h}{e}$
\begin{equation}\label{e2}
\Delta(r)=
\left\{%
\begin{array}{ll}
    \Delta,  & r< R \\
    0, &  r>R \\
\end{array}%
\right., \qquad
	V(r)=
	\left\{%
	\begin{array}{ll}
		\hbar v_F V,  & r< R \\
    0, &  r>R \\
	\end{array}%
	\right., \qquad
	\vec A(r)=
	\left\{%
	\begin{array}{ll}
		\frac{h}{e}\frac{\phi}{2\pi r}\vec{e}_{\theta},   & r< R \\
    0, &  r>R \\
	\end{array}%
	\right.
\end{equation}
as  presented in  Figure \ref{f0}, with
$\vec{e}_{\theta}$ is the unit vector for the azimuthal angle, 
$\vec\sigma=(\sigma_x,\sigma_y, \sigma_z)$  are Pauli matrices and $\mathbb{I}$ is the $2 \times 2$ unit matrix.

We  carry out our work by considering
the polar coordinates $(r,\theta)$ 
such that the Hamiltonian \eqref{e1},
in system unit
$(\hbar=v_F=1)$, can be written as
\begin{equation}\label{e8}
H= \begin{pmatrix}
V_- & e^{-i\theta}(-i\frac{\partial}{\partial r}-\frac{1}{r}\frac{\partial}{\partial \theta}-i\frac{\phi}{r})
\\ e^{i\theta}(-i\frac{\partial}{\partial r}-\frac{1}{r}\frac{\partial}{\partial \theta}+i\frac{\phi}{r}) &  V_+
\end{pmatrix}
\end{equation}
where we have set the two potentials $V_{\pm}=V\pm \Delta$. 
In fact, \eqref{e8} generalizes our previous work \cite{Belouad2018}
to magnetic flux case and certainly will affect the system and offer
interesting results. To obtain the eigenspinors $\Psi_m(r,\te)$, we use the fact that
the Hamiltonian \eqref{e8} commutes with the total angular momentum $J_z=L_z+\sigma_z/2$. This commutation requires  the separability of  $\Psi_m(r,\te)$ into the radial $R^{\pm}(r)$ and angular $\psi^{\pm}(\theta)$ parts and then we have 
\begin{equation}\label{e6}
\Psi_m(r,\te)=\begin{pmatrix} R^+_m (r)\psi_m^+(\theta) \\ R^-_{m+1} (r)\psi_{m+1}^-(\theta)  \end{pmatrix}, \qquad m\in \mathbb{Z}
\end{equation}
such that
\begin{equation}\label{e7}
\psi^+(\theta)=\frac{e^{im\theta}}{\sqrt{2\pi}}\begin{pmatrix} 1  \\ 0  \end{pmatrix}, \qquad \psi^-(\theta)=\frac{e^{i(m+1)\theta}}{\sqrt{2\pi}}\begin{pmatrix} 0  \\ 1  \end{pmatrix}
\end{equation}
are eigenstates of $J_z$ associated  to  the eigenvalues $m+\frac{1}{2}$. These will be used in the next to give the full solutions of the energy spectrum.

Now we determine
the radial parts by   solving  the eigenvalue equation $H\Psi(r,\theta)=E\Psi(r,\theta)$ for two regions:
outside
$r>R$ and  inside $r\leq R$  of the quantum dot as shown in  Figure \ref{f0}. Indeed
for $r>R$, 
we obtain two equations for 
the  radial components $R_{m}^{+}$ and $R_{m+1}^{-}$  
\begin{eqnarray}
&& \left(\frac{\partial}{\partial r}+\frac{m+1}{r}\right)
R^-_{m+1}(r)=iER^+_m(r)\label{e9}\\
&&
\left(\frac{\partial}{\partial r}-\frac{m}{r}\right)
R^+_{m}(r)=iER^-_{m+1}(r)\label{e10}
\end{eqnarray}
which can be handled
by injecting \eqref{e10} into \eqref{e9} to derive
a second differential equation satisfied by  $R_m^+(r)$
\begin{equation}\label{e11}
\left(r^2\frac{\partial^2}{\partial r^2}+r\frac{\partial}{\partial r}+r^2E^2-m^2\right)R_m^+(r)=0
\end{equation}
showing that the solutions are
 Bessel functions $ J_m (Er)$ of type. Moreover, 
the wave function of the incident electron, propagating along
 $x$-direction ($x=r\cos\theta$), takes the form
\begin{equation}\label{e12}
\Psi_i(r,\theta)=\frac{e^{ikx}}{\sqrt{2}} \begin{pmatrix} 1  \\ 1  \end{pmatrix}
=\frac{1}{\sqrt{2}} \sum_m i^mJ_m(kr)e^{im\theta} \begin{pmatrix} 1  \\ 1  \end{pmatrix}.
\end{equation}
Using this
together with the eigenstates
 \eqref{e7}, we write the
incident spinor corresponding to the present system as
\begin{equation}\label{e13}
\Psi_i(r,\theta)=\frac{1}{\sqrt{2}} \sum_m i^{m+1}\left[-iJ_m(kr)e^{im\theta}\begin{pmatrix} 1  \\ 0  \end{pmatrix}+J_{m+1}(kr)
e^{i(m+1)\theta}\begin{pmatrix} 0  \\ 1  \end{pmatrix}\right]
\end{equation}
and  also the reflected one
\begin{equation}\label{e14}
\Psi_r(r,\theta)=\frac{1}{\sqrt{2}}
\sum_m i^{m+1}a_m\left[-iH^{(1)}_m(kr)e^{im\theta}\begin{pmatrix} 1  \\ 0  \end{pmatrix}+H^{(1)}_{m+1}(kr)e^{i(m+1)\theta}\begin{pmatrix} 0  \\ 1  \end{pmatrix}\right]
\end{equation}
in terms of 
  the Hankel functions $H^{(1)}_m (kr)$ of the first kind,  the scattering coefficients
  $a_m$
  and the
wave number  $k=\pm E$.

In the second region $r< R$, we have all external sources ($V,\Delta, \phi$) applied to graphene
and then
 the eigenvalue equation gives
\begin{eqnarray}
&&
\left(\frac{\partial}{\partial r}+\frac{m+1}{r}+\frac{\phi}{r}\right)R^-_{m+1}(r)=i\left(E-V_+\right)R^+_m(r)\label{e15}\\
&&
\left(\frac{\partial}{\partial r}-\frac{m}{r}-\frac{\phi}{r} \right)R^+_{m}(r)=i\left(E-V_-\right)R^-_{m+1}(r) \label{e16}.
\end{eqnarray}
Expressing  \eqref{e16} as
\begin{equation}\label{e17}
R^-_{m+1}(r)=-\frac{i}{E-V_-}
\left(\frac{\partial}{\partial r}-\frac{m}{r}-\frac{\phi}{r}\right)R^+_{m}(r)
\end{equation}
and replacing in
\eqref{e15} we get a 
differential equation for $R_m^+(r)$
\begin{equation}\label{e18}
\left(r^2\frac{\partial^2}{\partial r^2}+r\frac{\partial}{\partial r}+\eta^2r^2-(m+\phi)^2\right)R_m^+(r)=0
\end{equation}
by defining
the  parameter
$\eta^2=|(E-V_+)(E-V_-)-\Delta^2|$. The solution of \eqref{e18} can be worked out
to get the transmitted wave as
\begin{equation}\label{e19}
\Psi_t(r,\theta)=\frac{1}{\sqrt{2}}\sum_m i^{m+1}b_{m}\left[-iJ_{m+\phi}(\eta r)e^{i({m+\phi})\theta}\begin{pmatrix} 1  \\ 0  \end{pmatrix}+\mu
 J_{m+\phi+1}(\eta r)e^{i(m+\phi+1)\theta}\begin{pmatrix} 0  \\ 1  \end{pmatrix}\right]
\end{equation}
where we have set  $\mu= \pm \frac{\eta}{E-V_-}
$ and $b_m$ are the scattering coefficients. It is clearly seen that the last solution is strongly
depending on the magnetic flux $\phi$ together with the potential barrier  and energy gap.

The eigenvalue problem of the present system does not allow to explicitly determine the eigenenergies associated to the above eigenspinors. This opens door to look for other alternatives and then to overcome such situation
one may  use
 the boundary condition. For this, we much 
 the eigenspinors 
 \beq
 \psi_i(R)+\psi_r(R)=\psi_t(R)
 \eeq
 at interface
 $ r=R$ to obtain the set of equations
\begin{eqnarray}
 &&
J_m(kR)+a_mH_{m}^{(1)}(kR)=b_{m+\phi}J_{m+\phi}(\eta R) e^{i\phi\theta }\label{e20}
\\
&&
J_{m+1}(kR)+a_mH_{m+1}^{(1)}(kR)=\mu b_{m} J_{m+\phi+1}(\eta R) e^{i\phi\theta }\label{e21}.
\end{eqnarray}
After straightforward algebras, we derive the
scattering coefficients $a_m(\phi)$ and $ b_{m}(\phi)$ in terms  of
the magnetic flux $\phi$
\begin{eqnarray}
 &&
a_m(\phi)=\frac{-J_{m+\phi}(\eta R)J_{m+1}(kR)+\mu J_{m+\phi+1}(\eta R)J_m(kR)}{J_{m+\phi}(\eta R)H^{(1)}_{m+1}(kR)-\mu J_{m+\phi+1}(\eta R)H_m^{(1)}(kR)}\label{e22}
\\
&& b_{m}(\phi)=\frac{J_m(k R)H^{(1)}_{m+1}(kR)- J_{m+1}(k R)H^{(1)}_m(kR)}{J_{m+\phi}(\eta R)H^{(1)}_{m+1}(kR)-\mu J_{m+\phi+1}(\eta R)H_m^{(1)}(kR)}\label{e23}.
\end{eqnarray}
We notice that the derived results so for reduce to those obtained in our previous work \cite{Belouad2018} by requiring a null magnetic flux.

\section{Current density and  cross section}

We will show how the current density and  cross section associated
to the present system
will be affected by  application of a magnetic flux $\phi$, which 
 will be done using the scattering coefficients $a_m$ and $b_m$ derived 
above.
Indeed, from the Hamiltonian \eqref{e1} we end up with the current density
\beq\lb{curr}
\vec{j} =\psi^\dagger\ {\vec\sigma} \psi
\eeq
where  $ \psi=\psi_i+\psi_r$ is the obtained eigenspinor
outside and $\psi=\psi_t$ is that inside the quantum dot of radius $R$. Considering the polar coordinates, we obtain 
the  radial component of \eqref{curr}
\begin{equation}\label{e24}
j_r=
\psi^\dagger \begin{pmatrix}
0 & \cos\theta-i~\sin\theta
\\ \cos\theta+i~\sin\theta &  0
\end{pmatrix}\psi.
\end{equation}
By focusing on
the reflected wave,  from \eqref{e24} we show that the corresponding radial current
can be written as
\begin{equation}\label{e25}
j^r_r(\phi)=\frac{1}{2}\sum_m A_{m,\phi}(kr)\begin{pmatrix}
0 & e^{-i\theta}
\\ e^{i\theta} &  0\end{pmatrix}\sum_m B_{m,\phi}(kr)
\end{equation}
where the two functions are given by
\begin{eqnarray}
A_{m,\phi}(kr) &=&(-i)^{m+1}\left[iH_m^{(1)*}(kr)\left(a_m^*(\phi)e^{-im\theta}\ \ \ a^*_{-m-1}(\phi)e^{im\theta}\right)\right.\nonumber \\
&&
\left.+
H_{m+1}^{(1)*}(kr)\left(a^*_{-m-1}(\phi)e^{i(m+1)\theta}\ \ \ a_m^*(\phi) e^{-(m+1)\theta}\right)\right]\label{e26}
\\
B_{m,\phi}(kr) &=& i^{m+1}\left[-iH_m^{(1)}(kr)\begin{pmatrix}a_m(\phi) e^{im\theta}  \\a_{-m-1}(\phi)e^{-im\theta}  \end{pmatrix}+H_{m+1}^{(1)}(kr)\begin{pmatrix}a_{-m-1}(\phi) e^{-i(m+1)\theta}  \\a_m(\phi) e^{(m+1)\theta} \end{pmatrix}\right]\label{e27}.
\end{eqnarray}
The current density \eqref{e25} can be worked out by
requiring the
 condition $ kr\gg1$. In this case,
the asymptotic behavior of the Hankel functions
is approximated by
\begin{equation}\label{e28}
H_m(kr)\simeq \sqrt{\frac{2}{k\pi r}}e^{i\left[kr-(2m +1)\frac{\pi}{4}\right]}
\end{equation}
and it can be injected
into
(\ref{e26}-\ref{e27}) to reduce
\eqref{e25} to the following
\begin{equation}\label{e29}
j^r_r(\theta,\phi)=\frac{4}{k\pi r}\sum^{m=\infty}_{m=0}|c_m(\phi)|^2[\cos\left(2m+1\right)\theta+1]
\end{equation}
where the involved coefficients are
\beq
|c_m(\phi)|^2=\frac{1}{2}\left(|a_m(\phi)|^2+|a_{-(m+1)}(\phi)|^2\right).
\eeq
showing that the derived current density is strongly
depending on  the magnetic flux $\phi$. Certainly, it will allow us to find new results and emphasis what makes difference
with respect to our previous work \cite{Belouad2018}.

Having obtained the radial current for the reflected wave,
we now
 determine the scattering cross section relative to the present system. It
 is given by the total reflected flux through a concentric circle $I_r^r$
divided by the incident flux per unit area $I^i/A_u$ \cite{C.Schulz2015M}
\beq
 \si=\frac{I_r^r}{I^i/A_u}
 \eeq
where $I_r^r$ can be calculated by integrating the radial reflected current over
the circle of angle $\theta$ to end up with a result in terms of the magnetic flux $\phi$
\begin{equation}
I^{r}_{r}(\phi)=\int_0^{2\pi} j^{r}_{r}(\theta,\phi)r d\theta=
\frac{8}{k}\sum^{m=\infty}_{m=0}|c_m(\phi)|^2
\end{equation}
while  for the incident wave $\psi_i(x)=\frac{e^{ikx}}{\sqrt{2}}\begin{pmatrix}
1\\
1\\
\end{pmatrix}$, we have $I^i/A_u=1$. Then it follows that  the scattering cross section coincides with
the total reflected flux.
To  study  the scattering on the quantum dot with different radius, it is convenient to introduce  the scattering efficiency $Q$. It is defined as the ratio between the scattering cross section $\sigma$ and the geometric one $kR$,
such as
\begin{equation}\label{e31}
Q(\phi)= 
\frac{4}{kR}
\sum_{m=0}^{m=\infty}|c_m(\phi)|^2.
\end{equation}
Once
the radial current and scattering efficiency are derived, we proceed next to numerically compute these quantities in terms of the involved physical parameters under suitable conditions. This will help us to understand how the magnetic flux will affect
the scattering on the quantum dot of radius $R$.

\section{Numerical results }
We numerically
study diffusion of Dirac fermions
 in a circular electrostatically
defined quantum dot of radius $R$ in gapped graphene
subject to the magnetic flux $\phi$ measured in quantum unit
$\phi_0=\frac{h}{e}$. To give a better understanding, we divide the incident electronic energy $E$ into
three regimes
indexed by $E<V_-$, $V_-<E<V_+$ and $V_+<E$. Then we will be interested to
the numerical results
of the scattering efficiency $Q$, coefficients $|c_m(\phi)|^2$ and far-field scattered current $j^r_r (\te, \phi)$ by choosing different configurations  of the involved physical parameters.

\begin{figure}[!hbt]
	\centering
\includegraphics[scale=0.25]{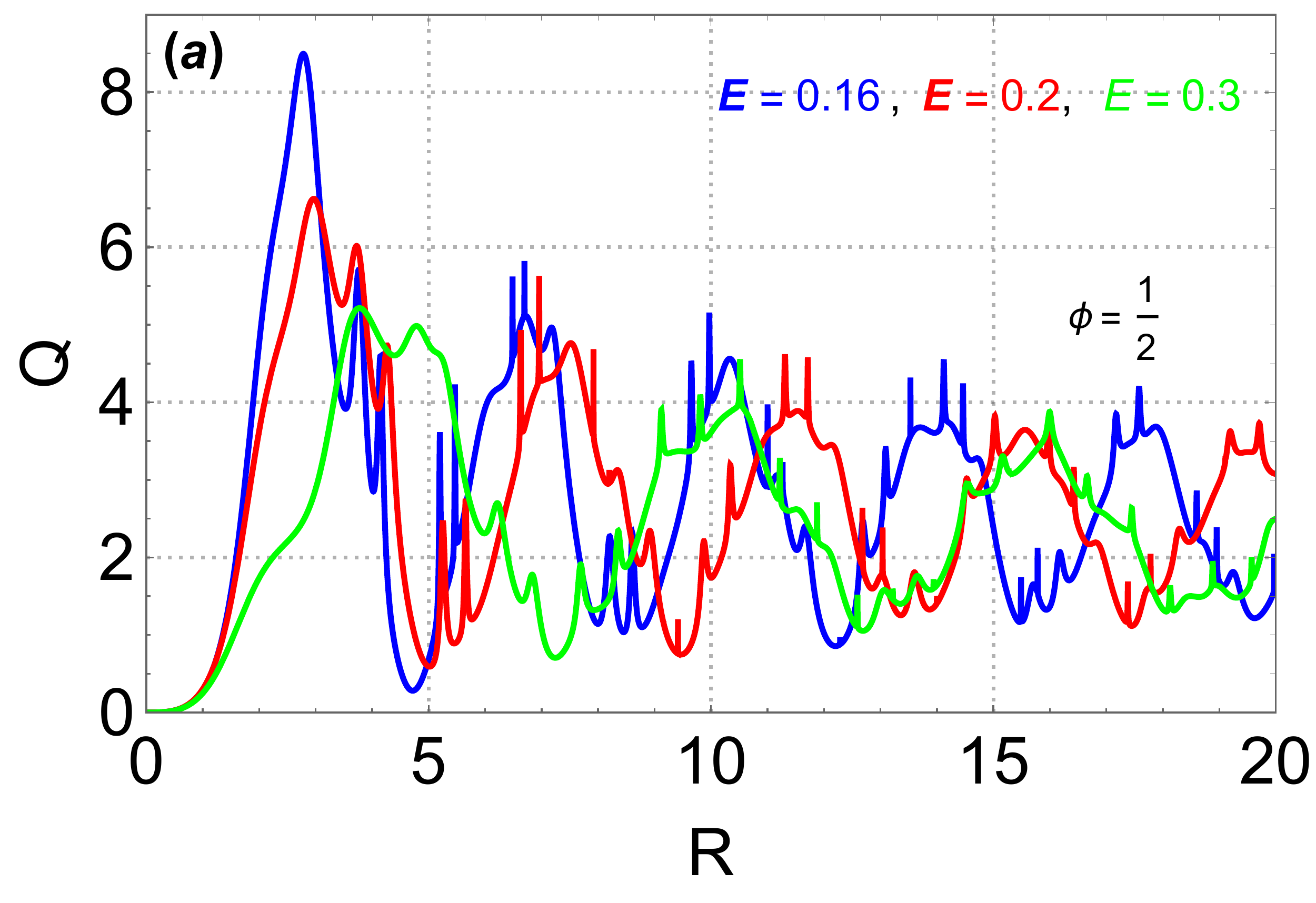}\ \ \ \ \ \ \ \ \includegraphics[scale=0.25]{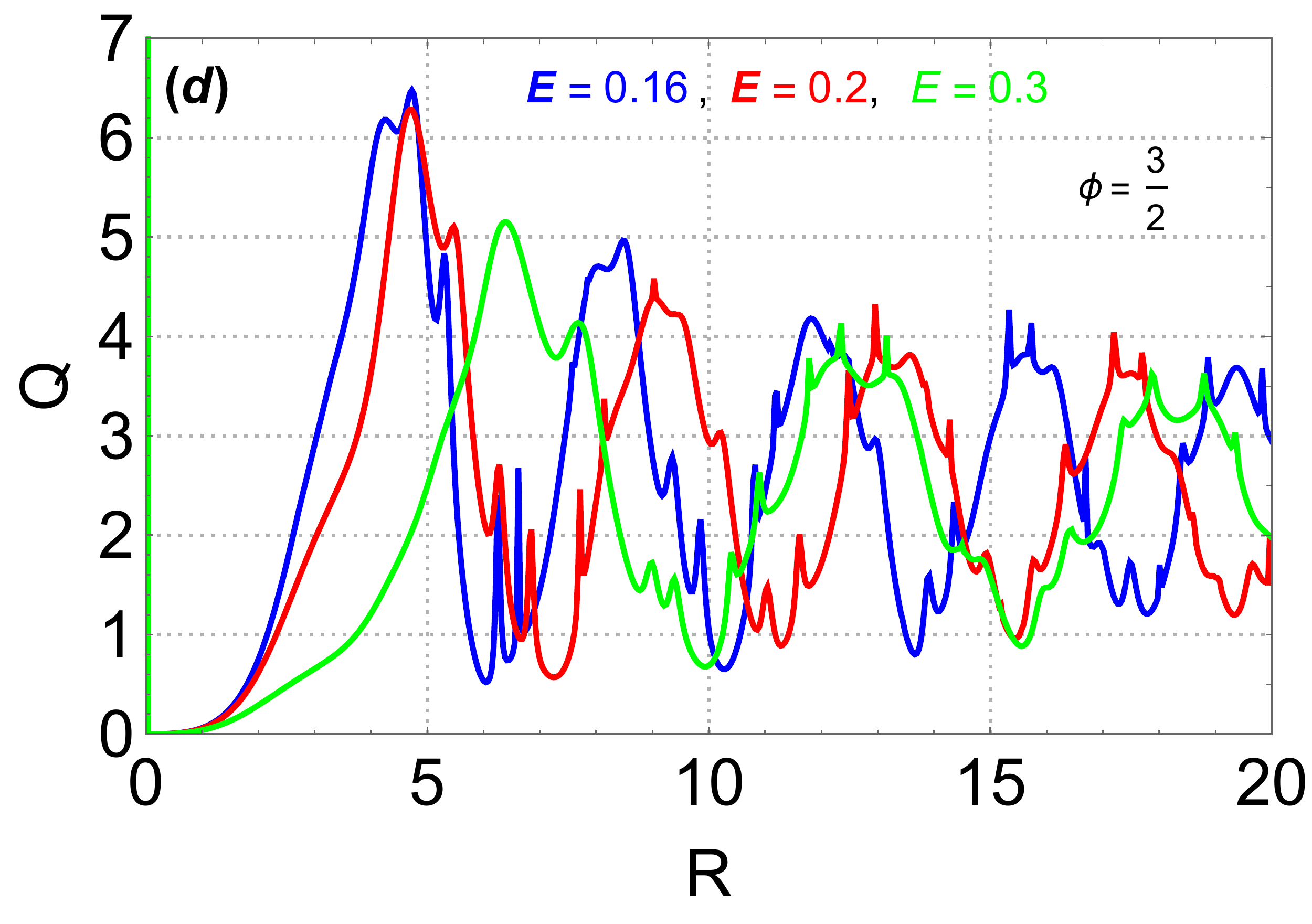}\\
\includegraphics[scale=0.25]{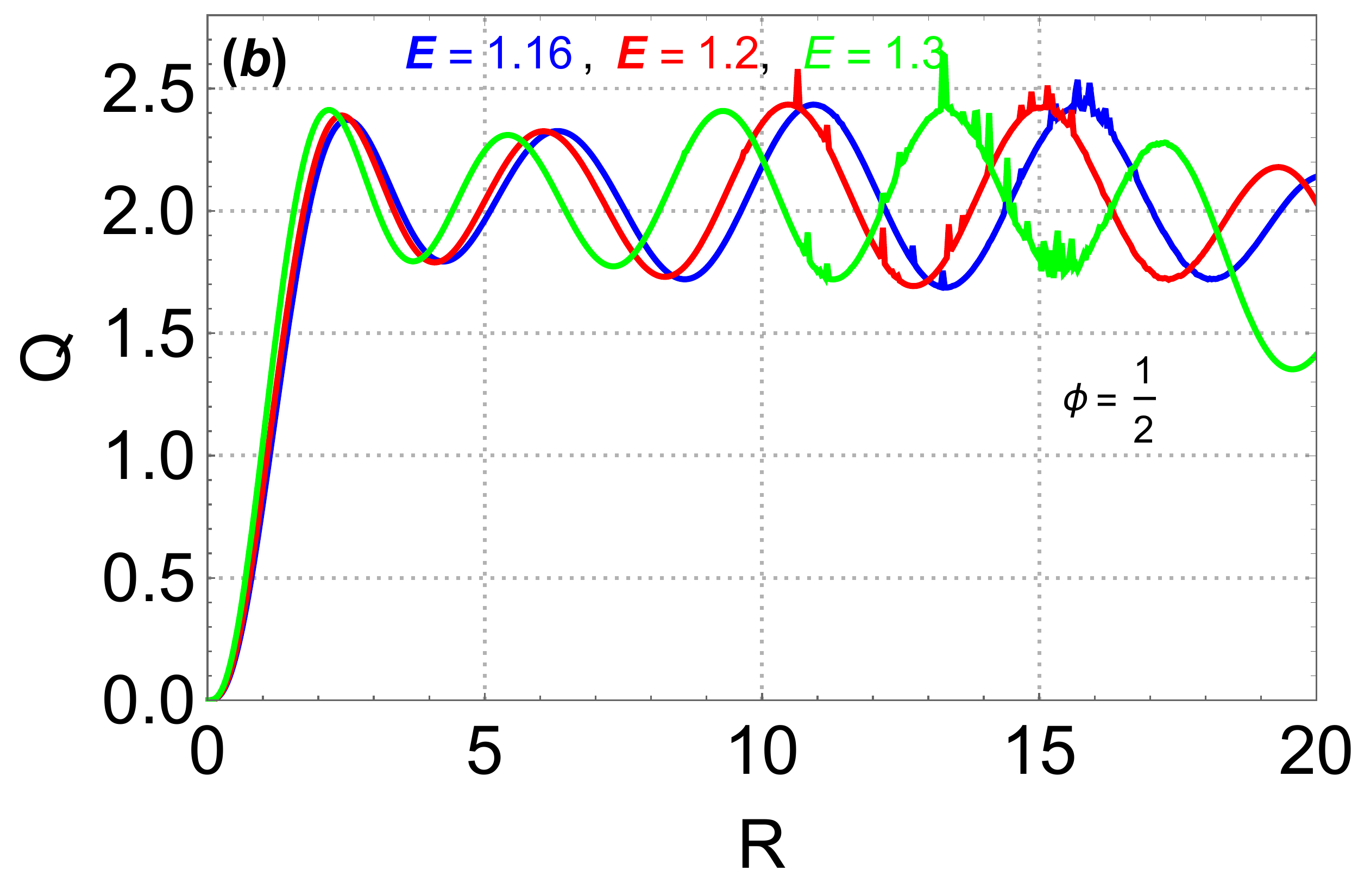}\ \ \ \ \ \ \ \ \includegraphics[scale=0.25]{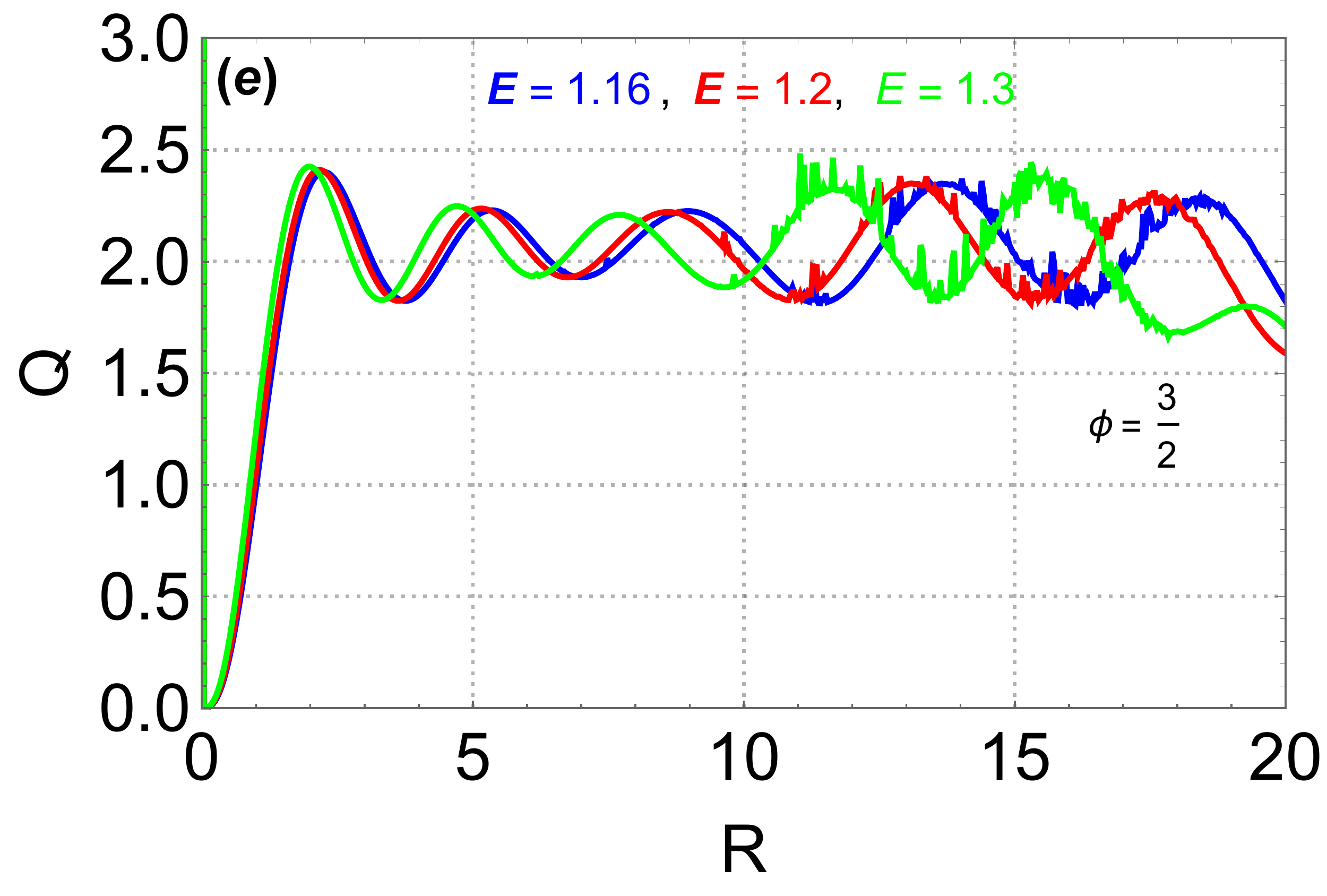}\\
\includegraphics[scale=0.25]{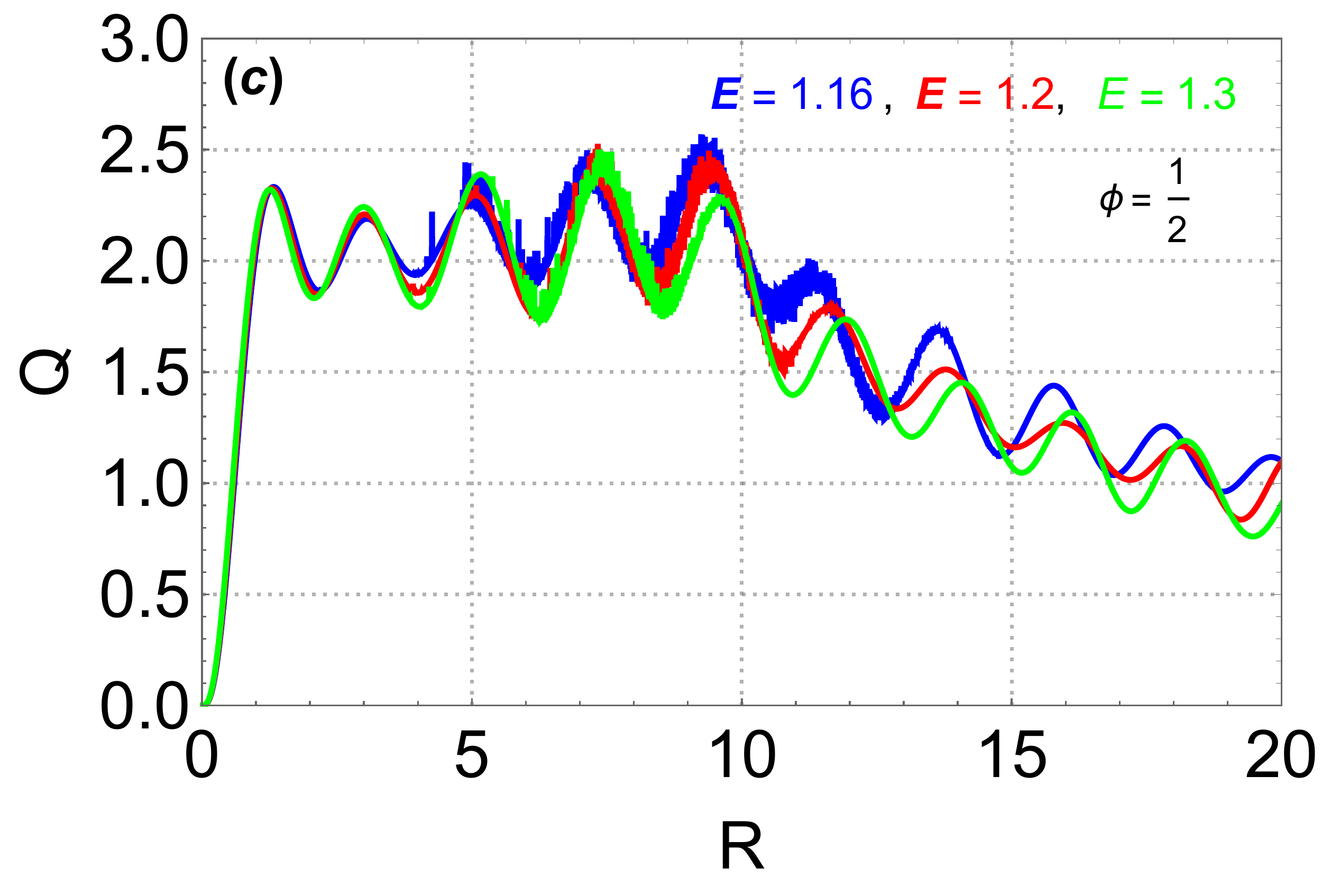}\ \ \ \ \ \ \ \ \includegraphics[scale=0.25]{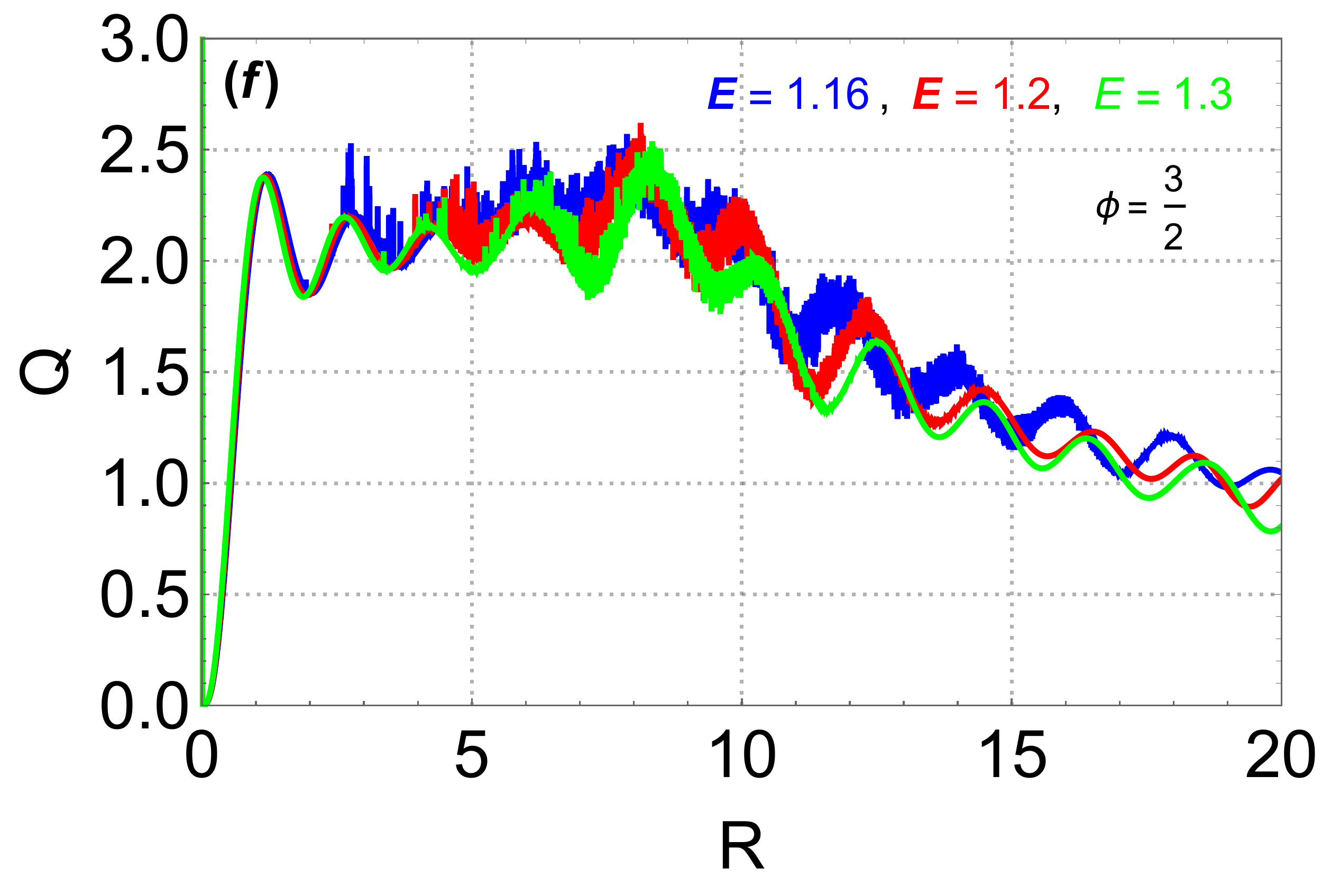}
\caption{\sf (color online) The scattering efficiency $Q$ as a function of the radius $R$ of the  quantum dot for different values of the incident energy $E$ with the  potential $V=1.3$, gap $\Delta=0.5$. Three regions are  (a) and (d): $E<V_-$, (b) and (e): $V_-<E<V_+$, (c) and (f): $V_+<E$.  The left panels correspond to the flux $\phi=\frac{1}{2}$ and  the right ones are for $\phi=\frac{3}{2}$.}\label{f1}
\end{figure}

Figure \ref{f1} presents the scattering efficiency $Q$ as a function of radius $R$
 for  $V=1.3$, $\Delta=0.5$, $\phi=\frac{1}{2},\frac{3}{2}$ and different values of the incident energy $E$. The energy regimes ($E<V_-,V_-<E<V_+,V_+<E$)  considered in the left and right panels are showing different behaviors. Indeed, in Figure \ref{f1}(a,d) we observe that
the electronic wave occupies the outside and  inside of the states of the quantum dot in the valence band. $Q$ is null in  the interval $0\leq R \leq 1$ for $\phi=\frac{1}{2}$ (Figure \ref{f1}a), while for $\phi=\frac{3}{2}$ (Figure \ref{f1}d) $Q$ has a maximum in the form of a peak when $R\rightarrow 0$.
As long as  $R$ increases,
$Q$ shows a damped oscillatory behavior with appearance of the emerging peaks and decrease of their amplitudes  for both  values of $\phi$. In addition, for  low energy the  amplitudes of $Q$ have maxima at specific values of $R$ and minima in the opposite. In Figures \ref{f1}(b,e)  the peculiarity of the energy dispersion of graphene becomes apparent and  $Q$ varies depending on 
$R$, $\phi$ and $E$. In fact, for $R$ tends to $0$ the amplitude of $Q$ is zero for $\phi=\frac{1}{2}$ and presents a sharp peak for $\phi=\frac{3}{2}$,
 for $R$ tends to $2$ we have a maximum value $Q=2.4$. However, when the size of the quantum dot becomes large, $Q$ has an oscillatory behavior with the appearance of some peaks. Such behavior is similar to that obtained in
 \cite{{C.Schulz2015M}} with  the requirement $\Delta=\phi=0$. Note that, the amplitude of the oscillations decreases by increasing the incident energy $E$.
For $E>V_+$,
in Figure \ref{f1}(c,f) one sees that   the regime outside and inside the quantum dot have electrons in the conduction band.
We notice that wen $R$  tends to zero, the results are similar to those corresponding to the regime $V_-<E <V_+$ (Figure \ref{f1}(b,e)) but
by increasing $R$
the three curves are superimposed and increase linearly up to a specific value of $R$ and where $Q$ shows an rapidly damped oscillatory behavior.
 It is clearly seen that
  the increase in magnetic flux $\phi$ is accompanied by  appearance of sharp peaks.
Compared to our previous work \cite{Belouad2018} dealing with the same system but without
flux, the main difference can be summarized  as follows. We have
almost the same oscillatory behavior but more  peaks in Figures \ref{f1}(a,d) for $E<V_-$,
 appearance of the peaks with  decrease in width of the oscillations
in Figures \ref{f1}(b, e) for $V_-<E<V_+$,
 peaks appear and  rapid damping of  $Q$ for $R> 10$ nm in Figures  \ref{f1}(c,f).
 These results show that the present system can be controlled under the appropriate
 choices of the flux $\phi$.\\

\begin{figure}[!hbt]
\centering
\includegraphics[scale=0.25]{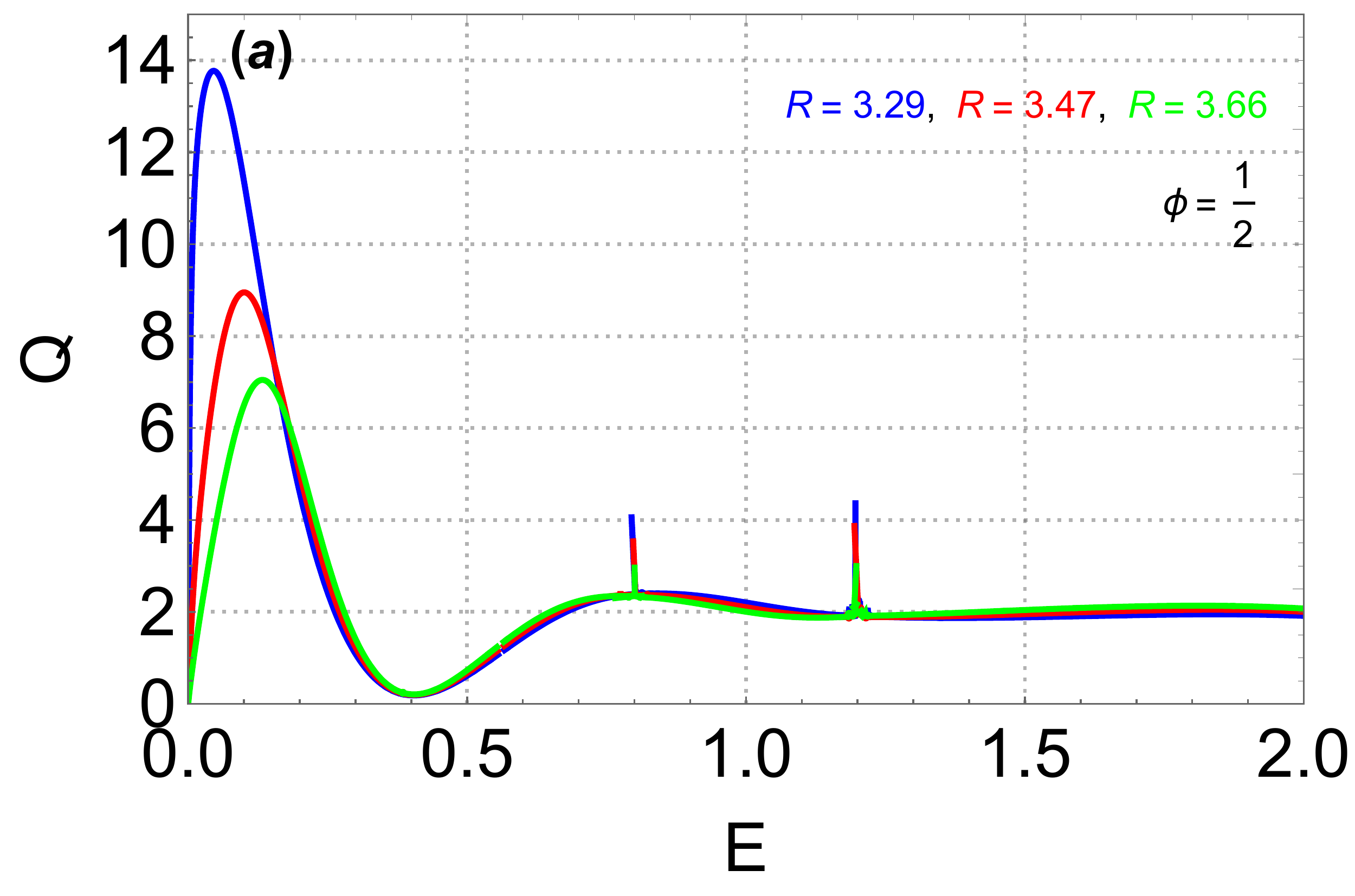}\ \ \ \ \ \ \ \ \includegraphics[scale=0.25]{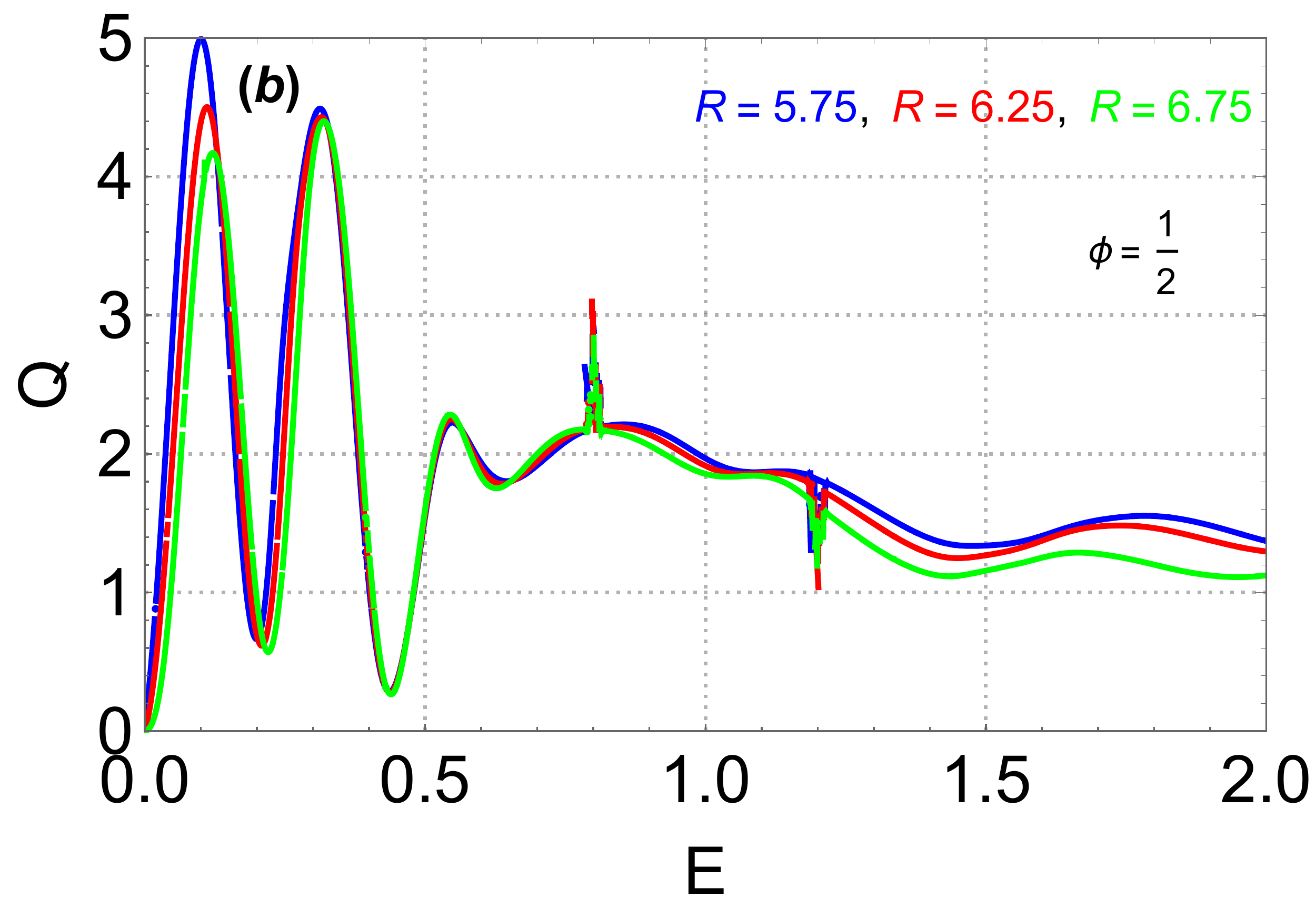}\\
\includegraphics[scale=0.25]{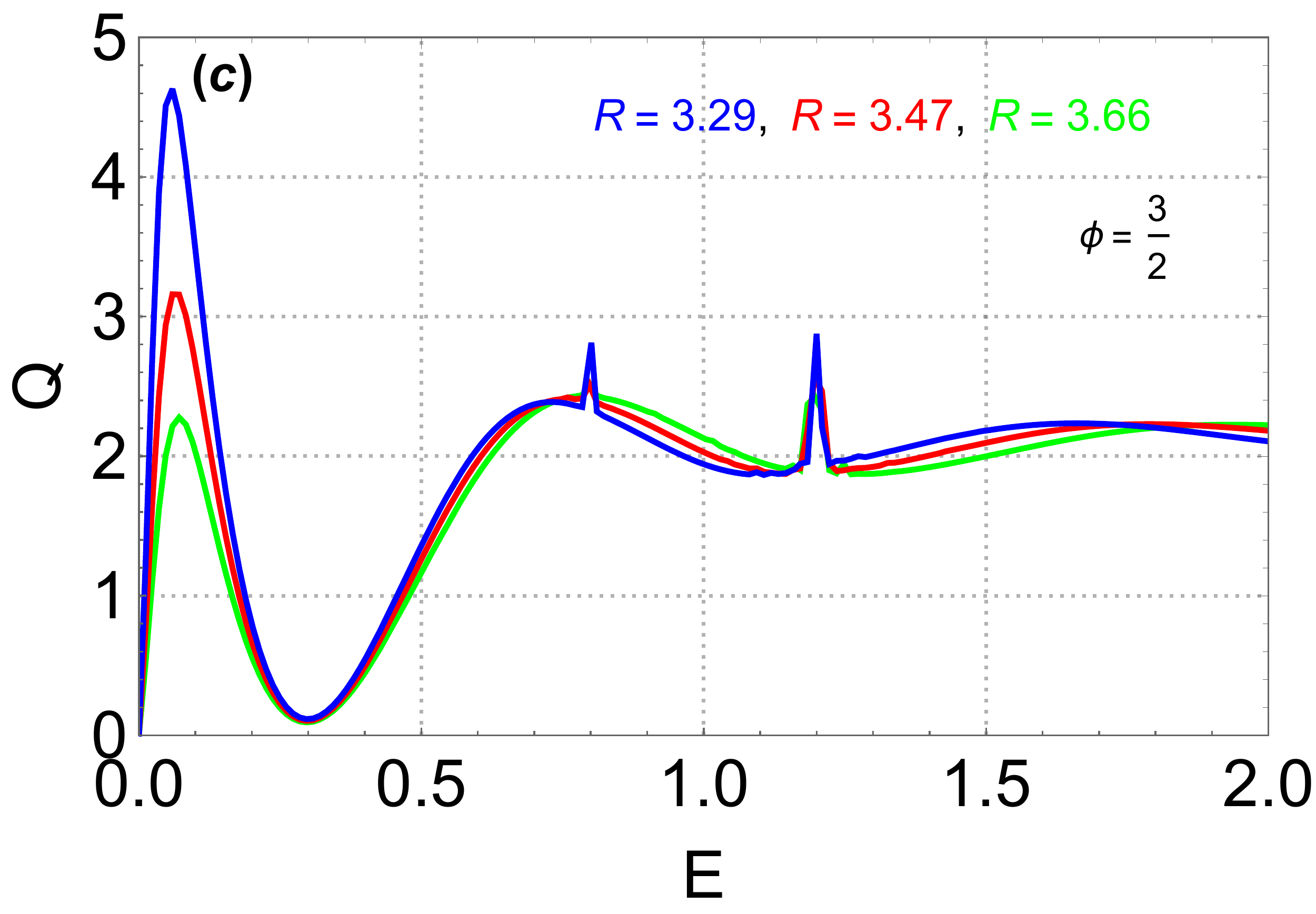}\ \ \ \ \ \ \ \ \includegraphics[scale=0.25]{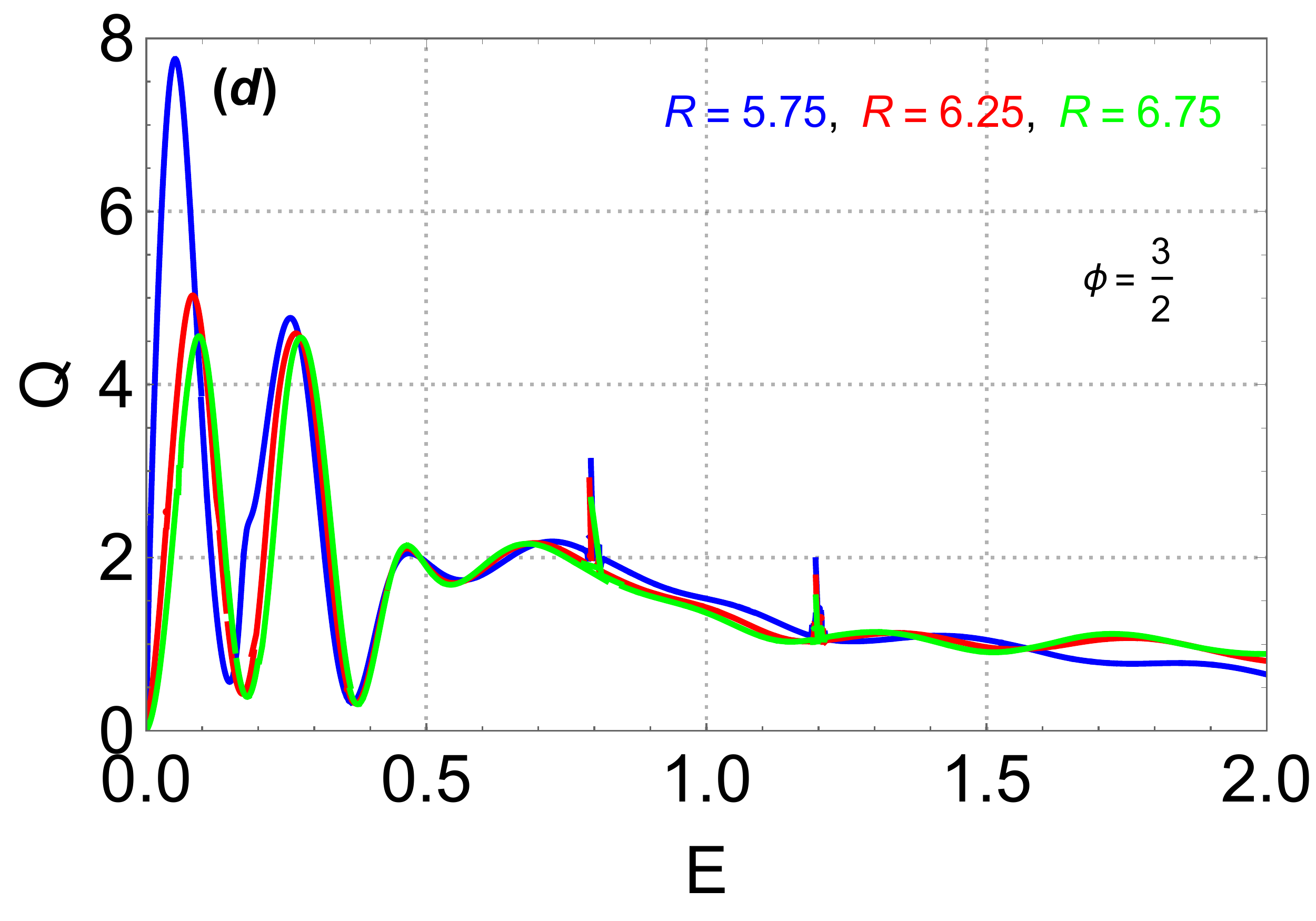}
	\caption{\sf (color online) {The scattering efficiency $Q$ as function of the incident energy $E$  for three values of the quantum dot radius $R$ with $\Delta = 0.2$ and $ V = 1$. (a,c): $R=3.29, 3.47, 3.66$, (b,d): $R = 7.30 ,7.5, 7.8$. The Upper panels correspond to the flux $\phi=\frac{1}{2}$ and  the Bottom ones are for $\phi=\frac{3}{2}$.} }\label{f2}
\end{figure}

We plot in Figure \ref{f2} the scattering efficiency $Q$ as a function of the incident energy $E$ for  $V=1$, $\Delta=0.2$,  $\phi=\frac{1}{2}, \frac{3}{2}$ and  different  values of radius $R$. Indeed,
 in Figure \ref{f2}(a,c) we study the behavior of $Q$ quantum dots of small sizes $R$ ($3.29, 3.47, 3.66$) in three regimes. In regime  $E <V_-=0.8$, one observes that for $E\rightarrow 0$ we have a null $Q$ but as long as 
 $E$ increases  also $Q$ increases towards a maximum value $13.75$ for $\phi=\frac{1}{2}$ and $4.75$ for $\phi=\frac{3}{2}$. After that $Q$ decreases to a minimum value when $E$ increases towards $V_-$. 
 For the  regime $V_-<E<V_+$, we observe that $Q$ decreases linearly when $E$ increases. However in the regime $E>V_+=1.2$, $Q$ remains constant for $\phi=\frac{1}{2}$ and shows a small increase for $\phi=\frac{3}{2}$. Note that when $E=V_-=0.8$ and $E=V_+=1.2$ there is  appearance of the peaks due to the excitation of the normal modes of the quantum dot \cite{{C.Schulz2015M}}.
The case of  large quantum dot sizes $R$ is shown in Figures \ref{f2}(b,d) such that in  $E <V_-$, $Q$  exhibits an oscillatory behavior characterized by sharp and broad maxima 
for specific values of $E$.
For $V_-<E <V_+$, $Q$ has a low amplitude damped oscillatory behavior.
In the third regime $E>V_+$,
$Q$ decreases almost linearly for $\phi=\frac{3}{2}$ and of shows weak oscillation for $\phi=\frac{1}{2}$ when $ E $ increases. In addition, we observe  the existence of peaks at $V_-$ and $V_+$ energies when the incident electron passes from one regime to another.
We notice that, there is a difference in the oscillatory behavior of  $Q$
for the value $\phi= 3/2$ as shown
Figures \ref{f2}(b,d) and  plus more peaks at points of the resonances $E = V_-$ and
$E=V_+$.\\

\begin{figure}[!hbt]
	\centering
	\includegraphics[scale=0.25]{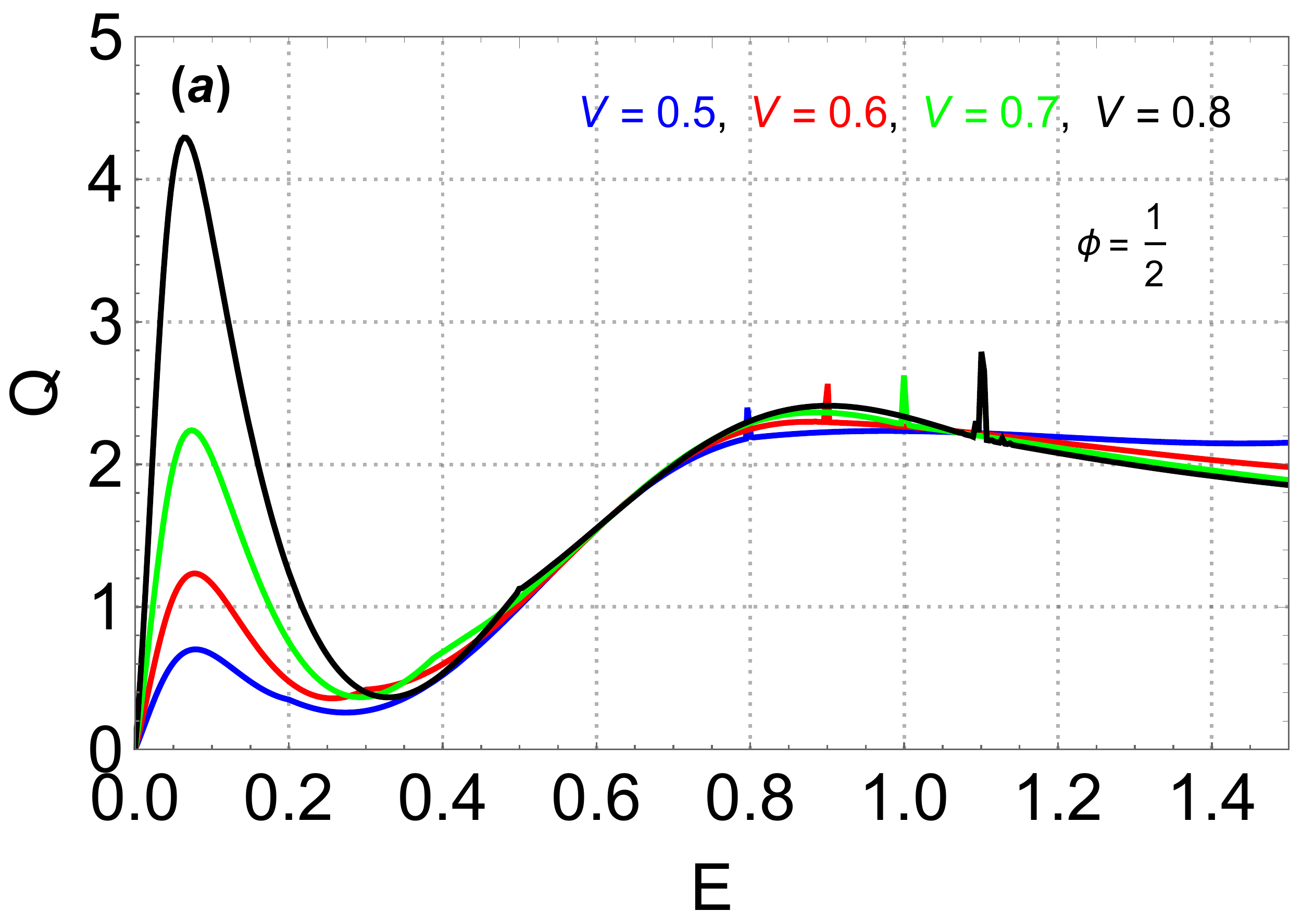}\ \ \ \ \ \ \ \  \includegraphics[scale=0.25]{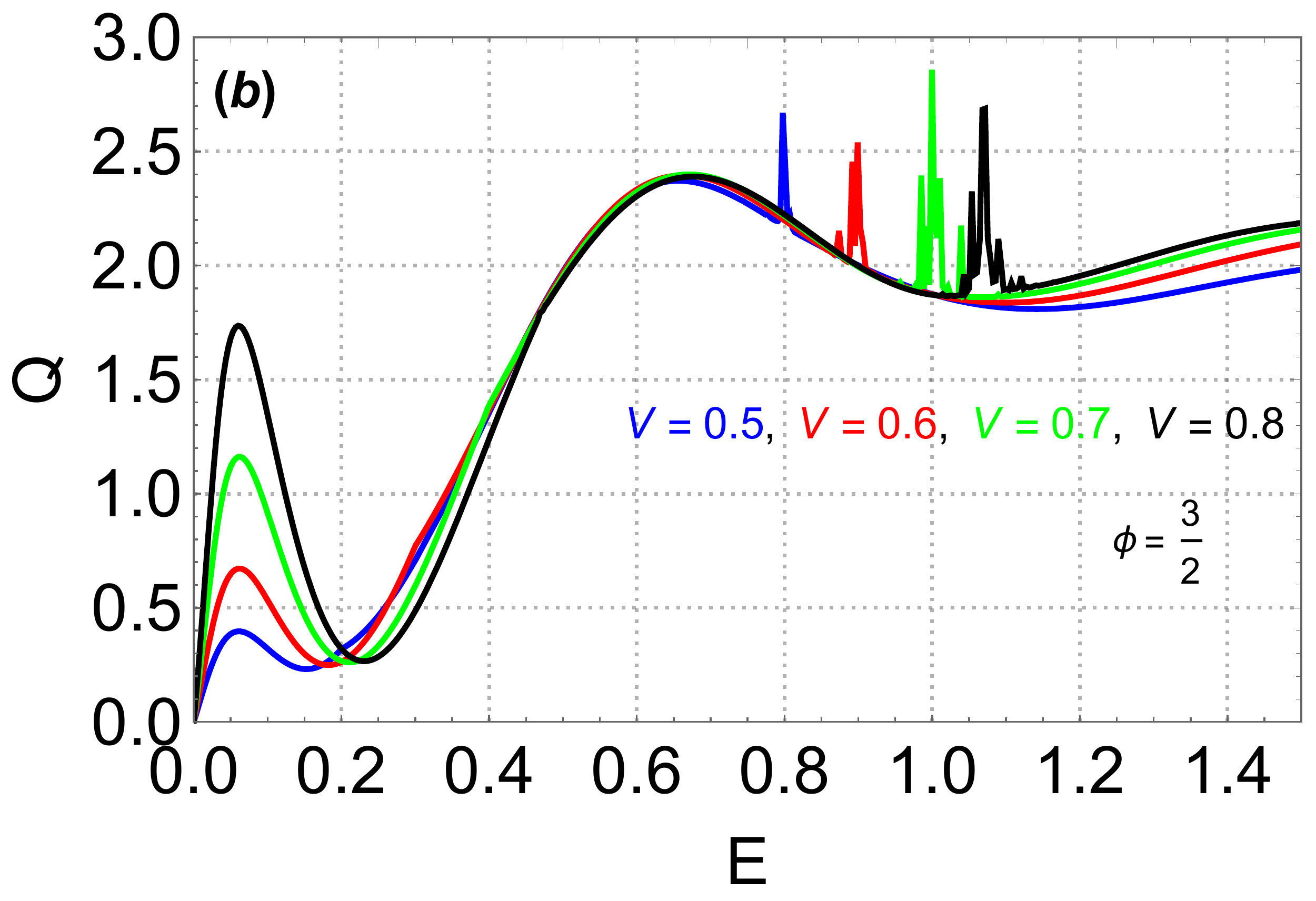}
	\caption{\sf(color online) The scattering efficiency $Q$ as function of the incident energy $E$ for five values of the potential height $V = 0.1, 0.2, 0.3, 0.4,0 .5$ with $\Delta = 0.3$ and $R=4$. (a): $\phi=\frac{1}{2}$ and
	(b): $\phi=\frac{3}{2}$. }\label{f4}.
\end{figure}

To underline the effect of the potential on the scattering phenomenon, we plot in Figure \ref{f4} the scattering efficiency $Q$ as a function of
the incident energy $E$ for five values of the potential $V$
with $\Delta=0.3$, $R=4$ where (a): $\phi=\frac{1}{2}$
and (b): $\phi=\frac{3}{2}$. We observe
for  $E=0$, $Q$ is zero whatever the value of $V$ and $\phi$. When $E$ increases all the curves show a maximum for each value of $V$, then $Q$ decreases to a minimum value depending on $V$ and
after that $Q$ increases by exhibiting some peaks. It is clearly seen that the amplitude of $Q$ increases as long as  $V$ increases, which is in agreement with literature \cite{H.2015Fehske}. As it is apparent from the Figure 4 (a,b) the increase in the magnetic flux $\phi$ is accompanied by the appearance of the peaks at the resonance point
$E = V_+$.\\

\begin{figure}[!hbt]
\centering
\includegraphics[scale=0.3]{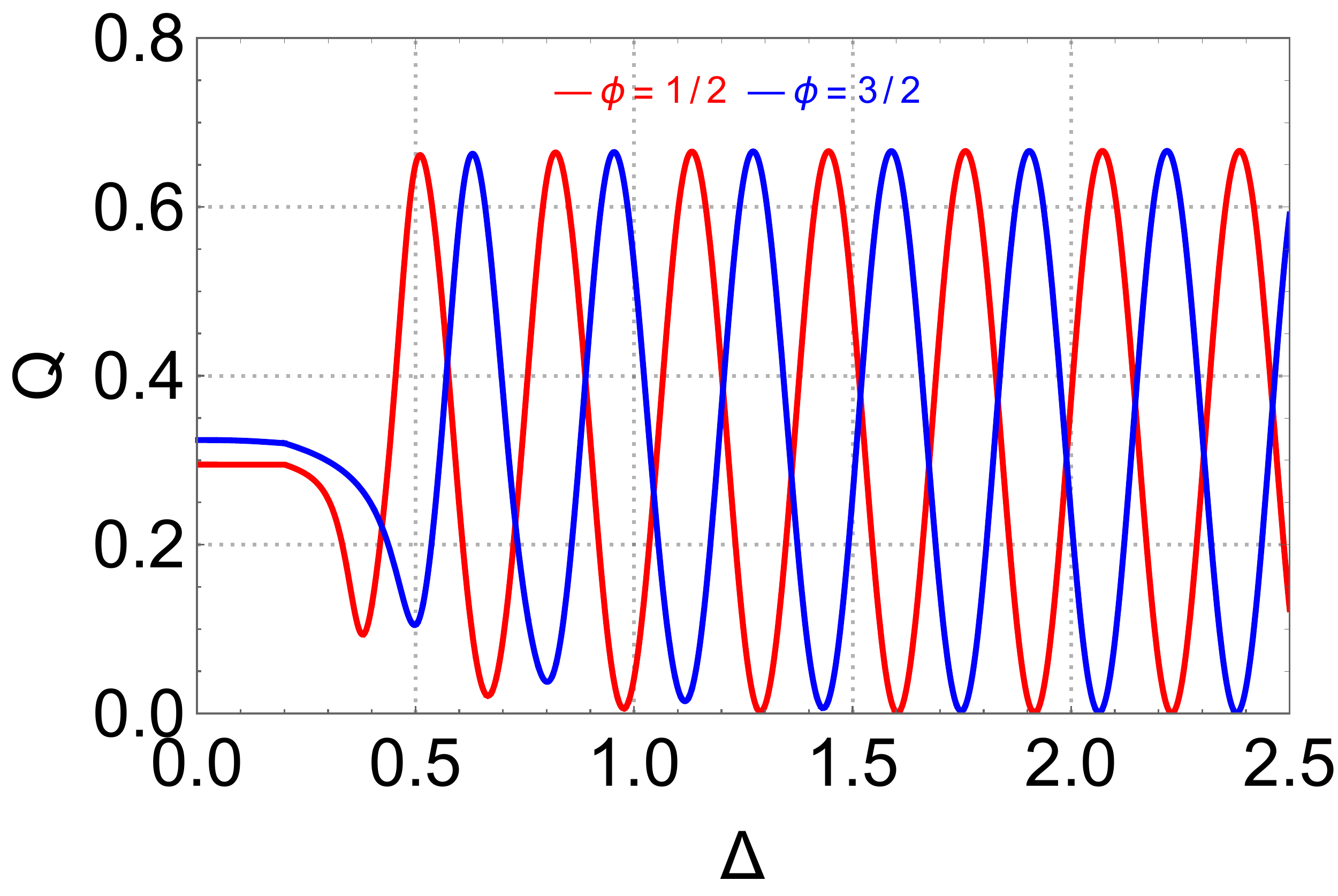}
\caption{\sf(color online) The scattering efficiency $Q$ as function of  the gap
$\Delta$ for  $E=1.2$, $V=1$, $R=10$,  $\phi=\frac{1}{2}$ (red line), $\phi=\frac{3}{2}$ (blue line).}\label{f6}
\end{figure}

Figure \ref{f6} presents 
the scattering efficiency $Q$ as a function of the gap $\Delta$ applied inside the quantum dots for $R=10$, $V=1$ and $E=1.2$
such that 
$\phi=\frac{1}{2}$ (red curve) and $\phi=\frac{3}{2}$ (blue
curve). We observe that when $\Delta\rightarrow 0$, $Q$ got fixed values with $Q=0.5$ for $\phi=\frac{1}{2}$ and $Q=0.27$ for $\phi=\frac{3}{2}$.  $Q$ decreases until a minimum value depending on $\phi$ and after that it shows a continuous oscillatory behavior for both values of  $\phi$ with $Q\left(\phi=\frac{1}{2}\right)\neq Q\left(\phi=\frac{3}{2}\right)$. Note that, the amplitudes of Q vary up to $\Delta=1$ and $\Delta=1.5$ for $\phi=\frac{1}{2}$ and $\phi=\frac{3}{2}$ respectively. After these values, their amplitudes remain constant for both values of the flux $\phi$.\\

\begin{figure}[!hbt]
\centering
\includegraphics[scale=0.25]{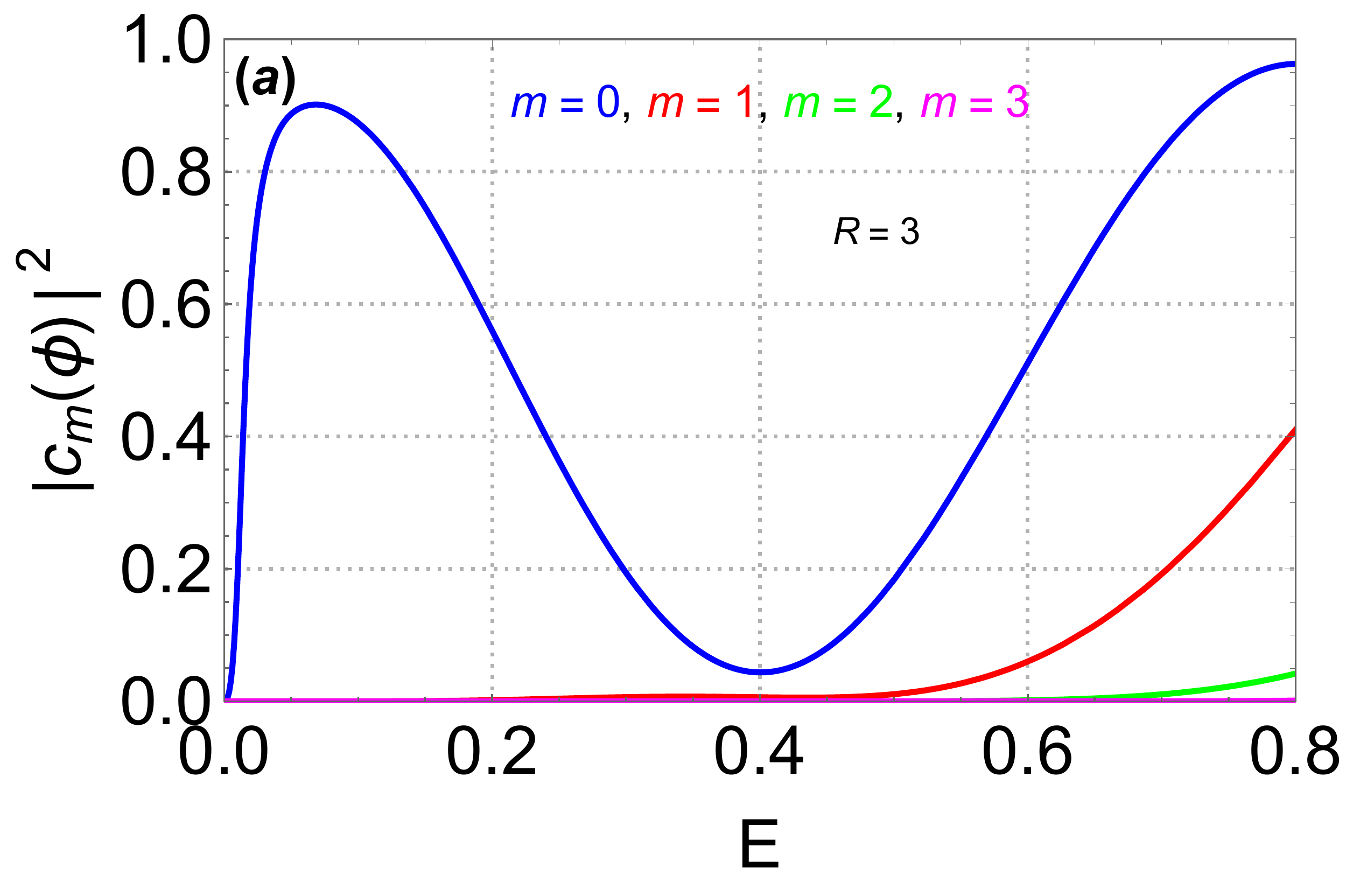}\ \ \ \ \ \ \ \ \includegraphics[scale=0.25]{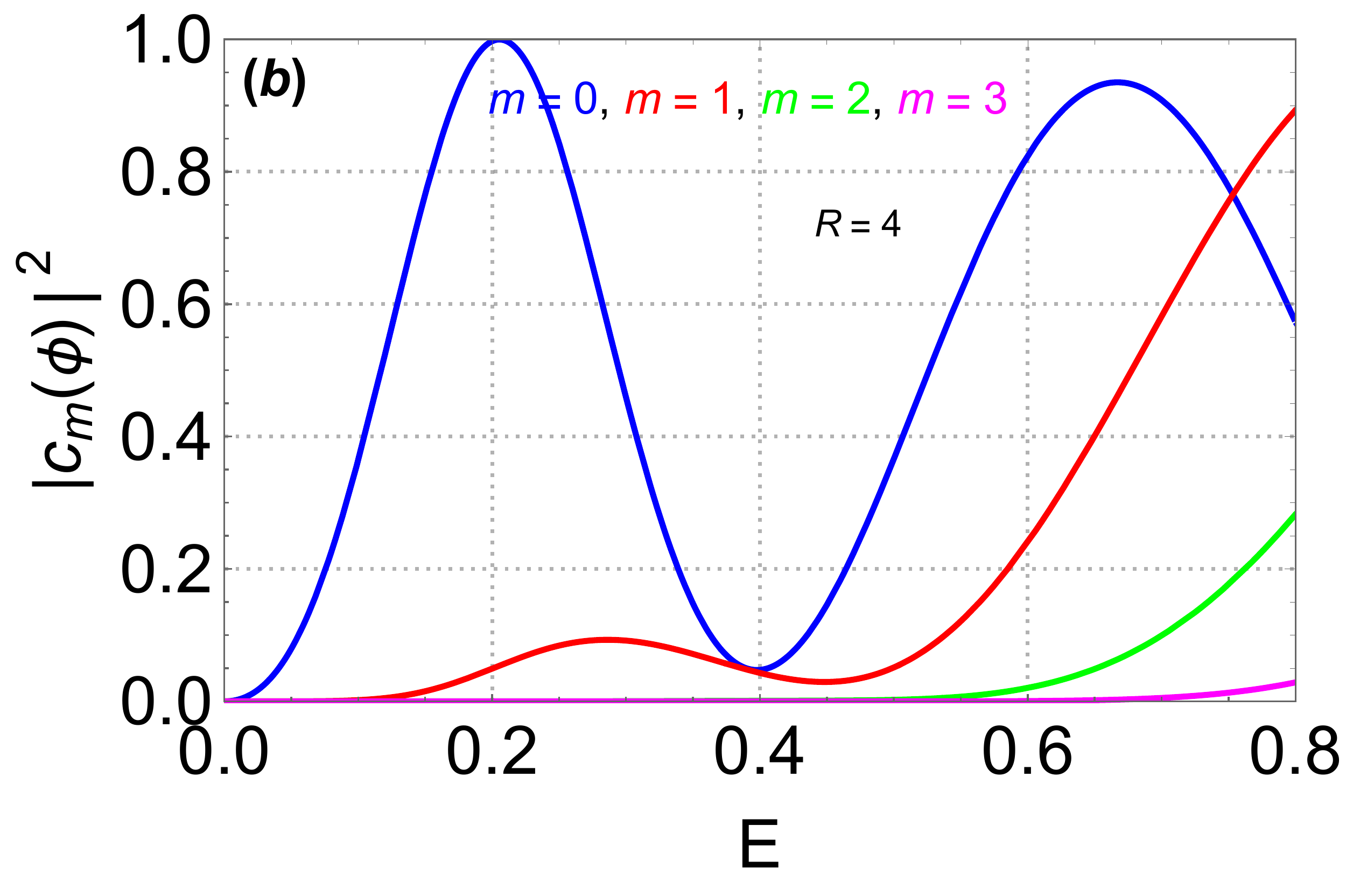}\\
 \includegraphics[scale=0.25]{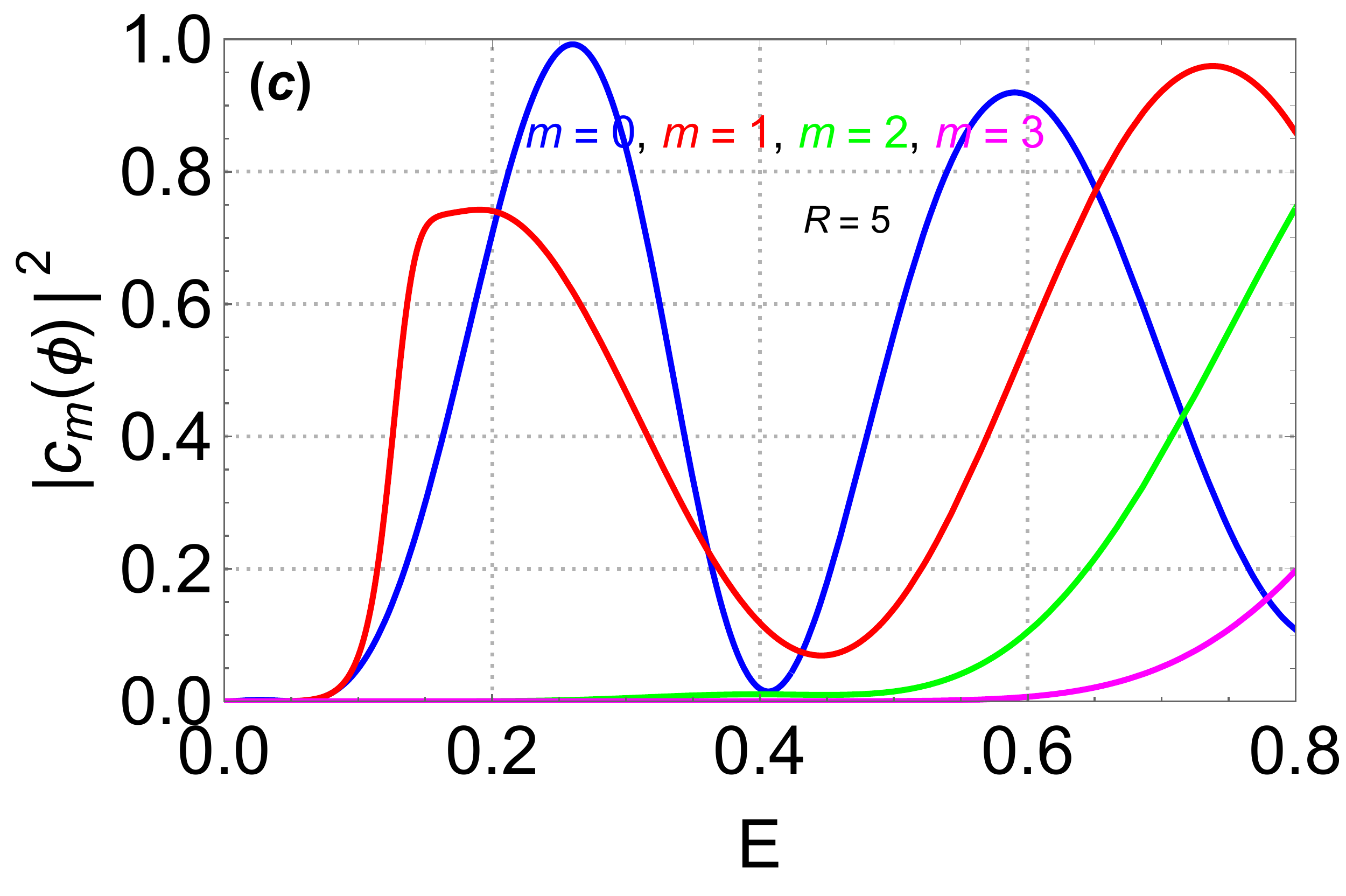}\ \ \ \ \ \ \ \ \includegraphics[scale=0.25]{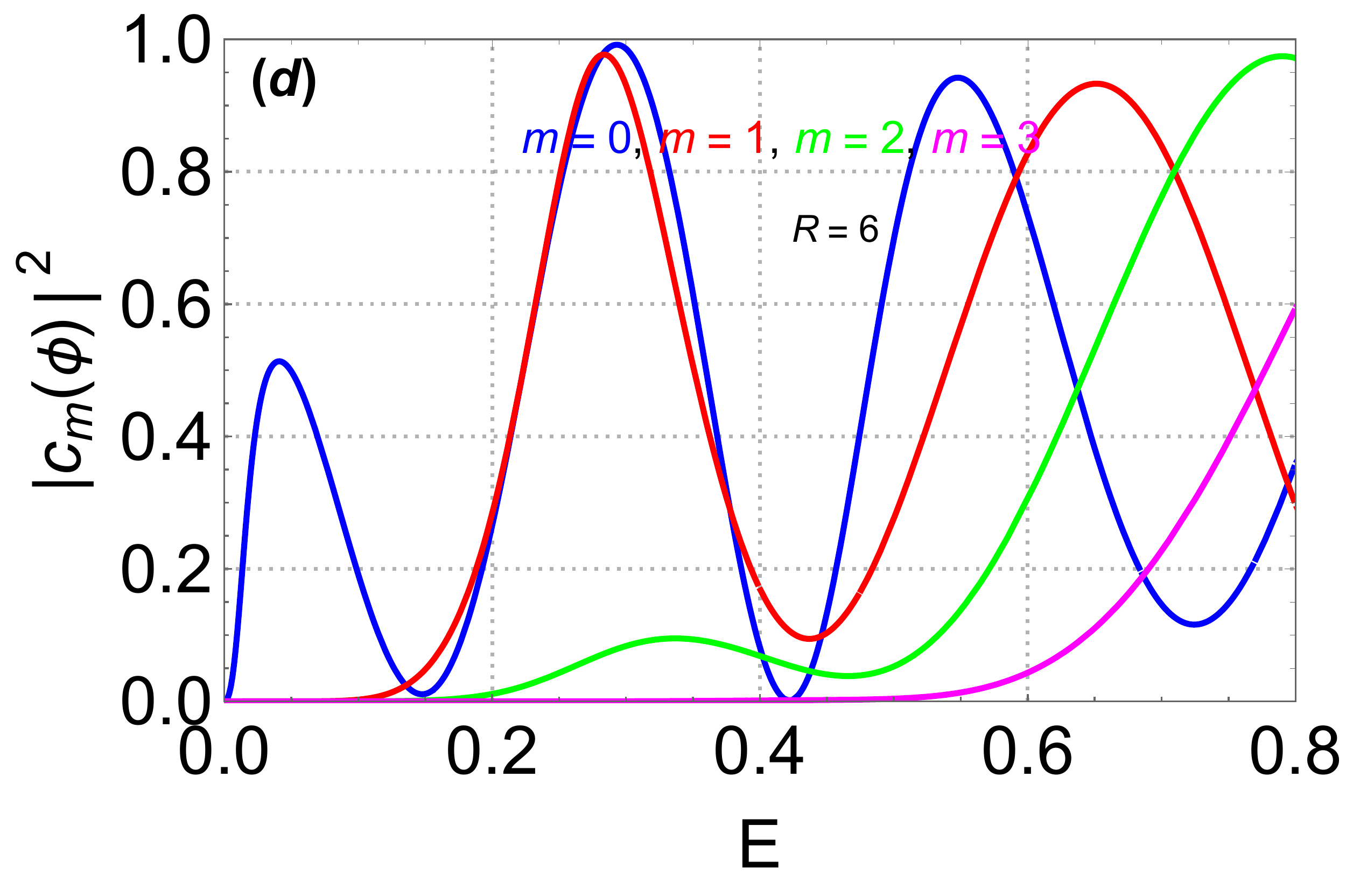} \\
 \includegraphics[scale=0.25]{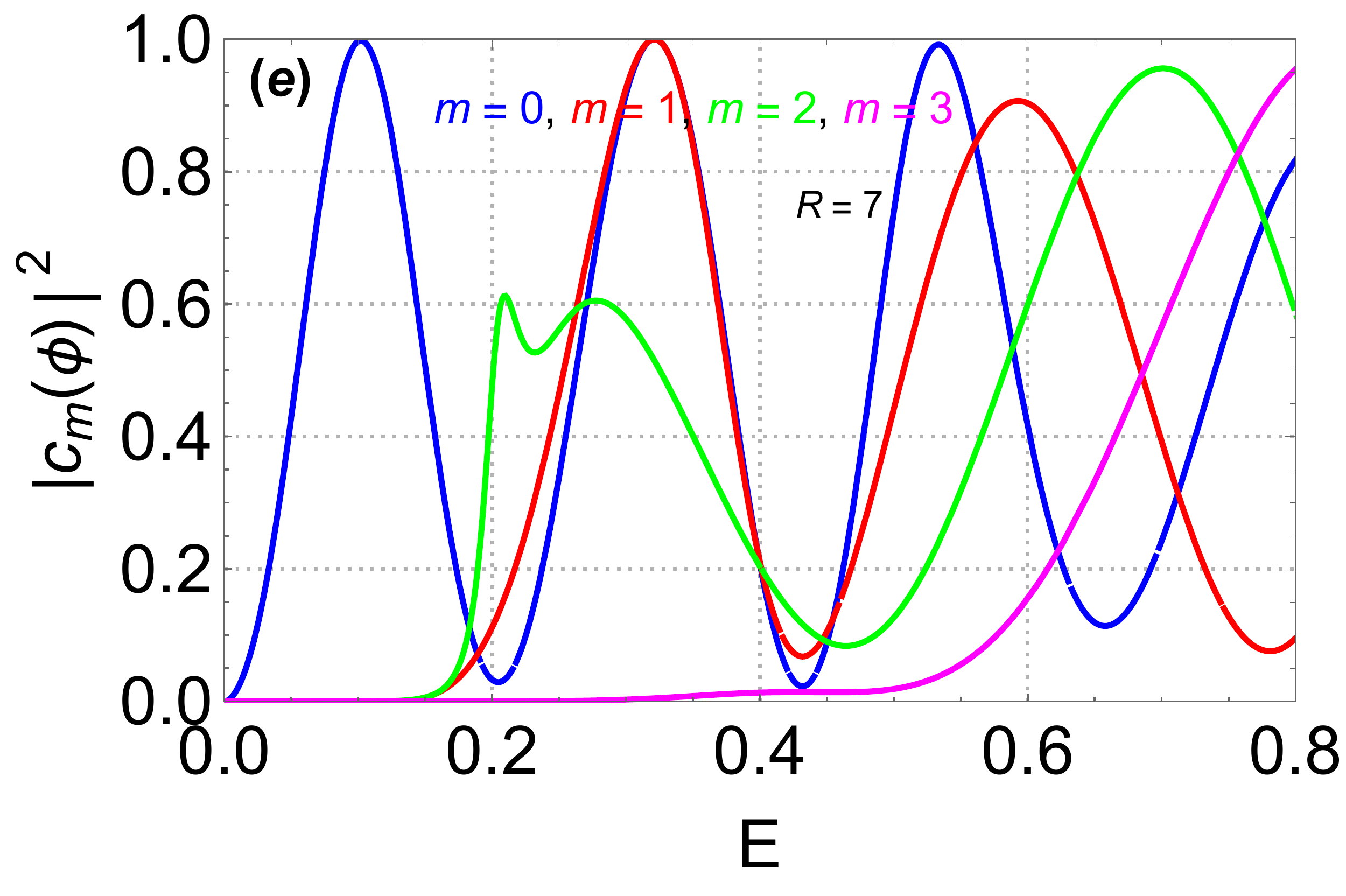}\ \ \ \ \ \ \ \ \includegraphics[scale=0.25]{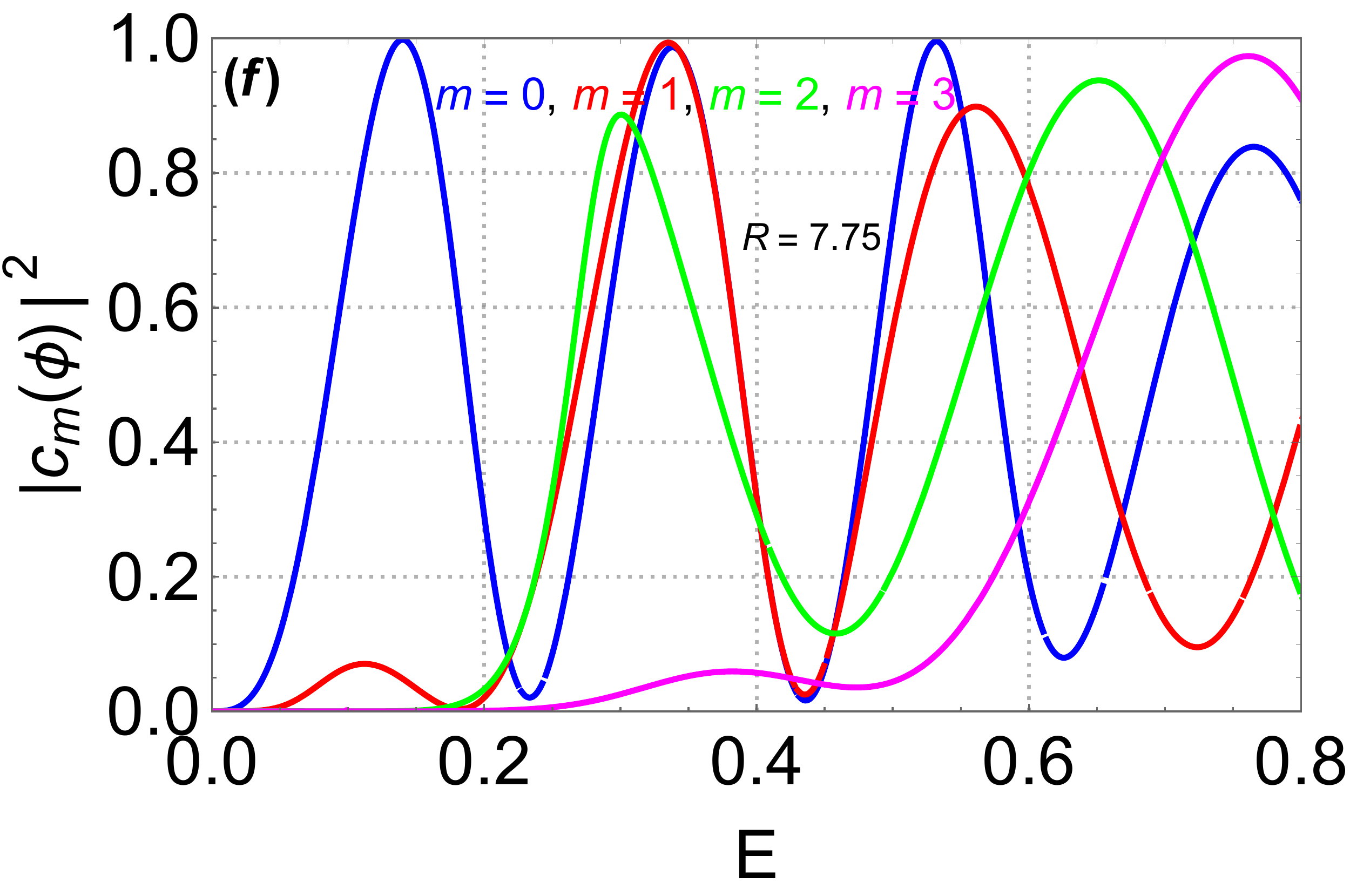}
	\caption{\sf(color online) The square modulus of the scattering coefficients $|c_m(\phi)|^2 $  as function of the energy $E$ for $m = 0,1,2,3$
	and for six values of the radius $R$ of the  quantum dot radius
	with $V=1$, $\Delta=0.2$ and flux $\phi=\frac{1}{2}$. (a): $R = 3$, (b): $R = 4$,
	(c): $R = 5$, (d): $R = 6$, (e): $R = 7$, (f):  $R = 7.75$. }\label{f7}
\end{figure}

To identify the resonances we plot in Figure \ref{f7} 
the square modulus of the scattering coefficients
$|c_m(\phi)|^2$  ($m=0, 1, 2, 3$) as function of the incident
energy $E$ for the choice $V = 1$, $\Delta=0.2$, $\phi=\frac{1}{2}$ such that  (a): $R=2$, (b):
$R=4$, (c): $R=5$, (d): $R=6$, (e): $R=7$, (f): $R=7.75$. It is clearly seen that close to $E=0$ all $c_m$ vanish except the lowest 
one $c_0$. By increasing $E$, the  contribution of
the scattering coefficients of higher orders $(m=1, 2,
3)$ starts to show up. For larger energy the $|c_m(\phi)|^2$ tend to an oscillatory behavior. We notice that the presence of an gap $\Delta$ and flux $\phi$ increases the number of oscillations. For not too large $E$,
 the successive onset of modes is interspersed with sudden and sharp peaks of different $|c_m(\phi)|^2$  \cite{C.Schulz2015, C.Schulz2015M}.
These resonances of normal modes of the quantum  dot lead to the sharp peaks in the scattering coefficient $Q$ found in Figure \ref{f2}.\\

\begin{figure}[!hbt]
	\centering
\includegraphics[scale=0.25]{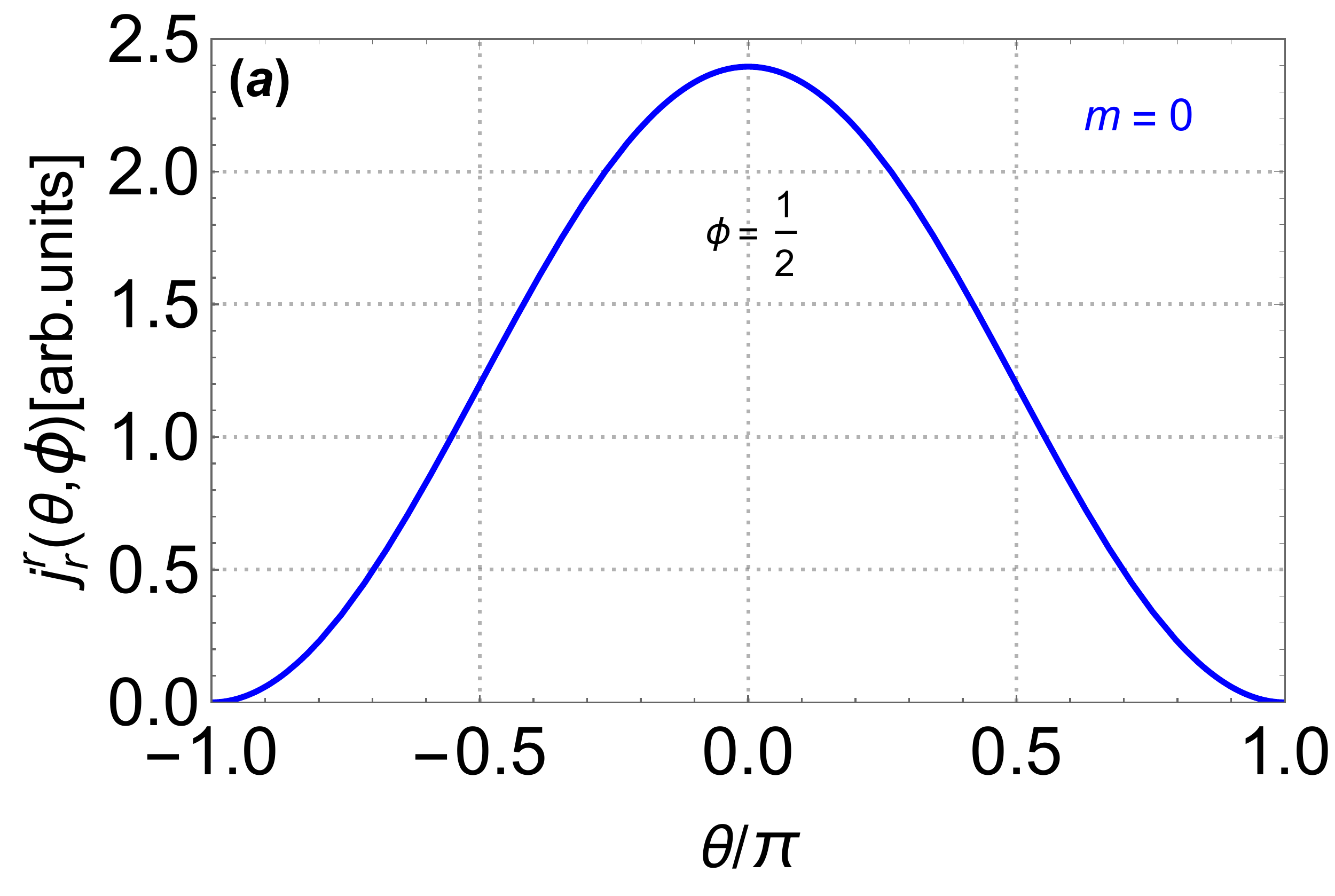}\ \ \ \ \ \ \ \ \includegraphics[scale=0.25]{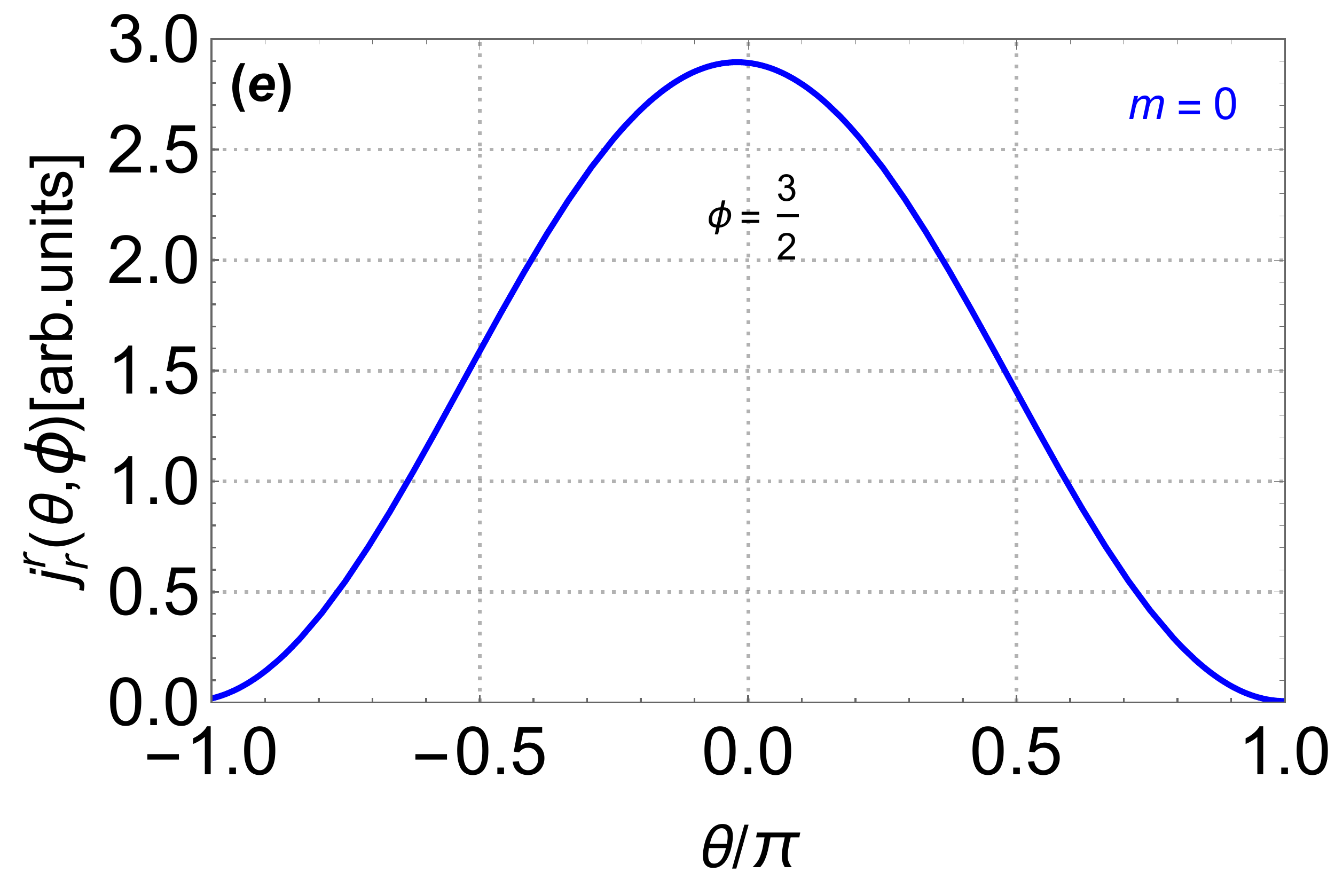} \\
\includegraphics[scale=0.25]{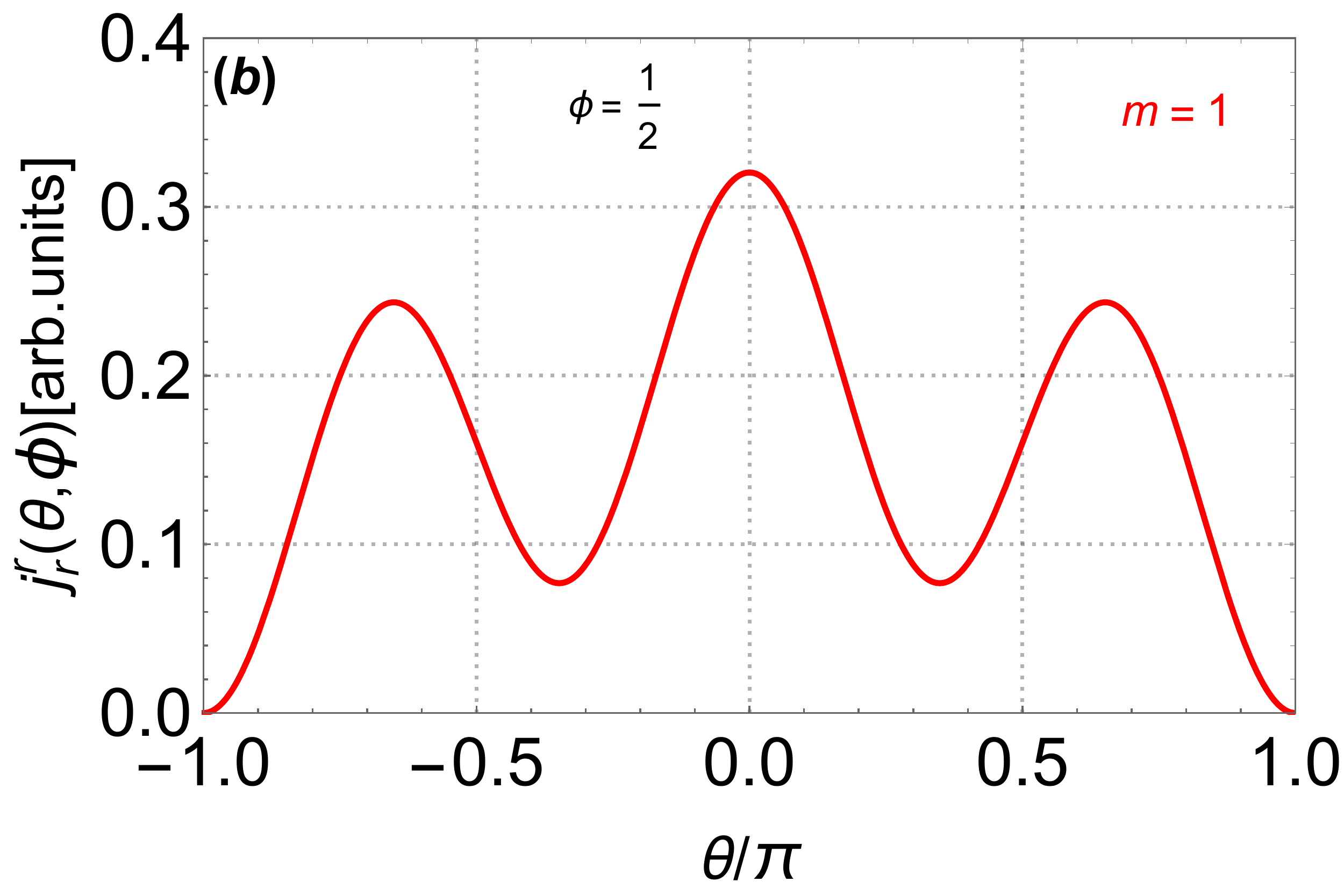}\ \ \ \ \ \ \ \  \includegraphics[scale=0.25]{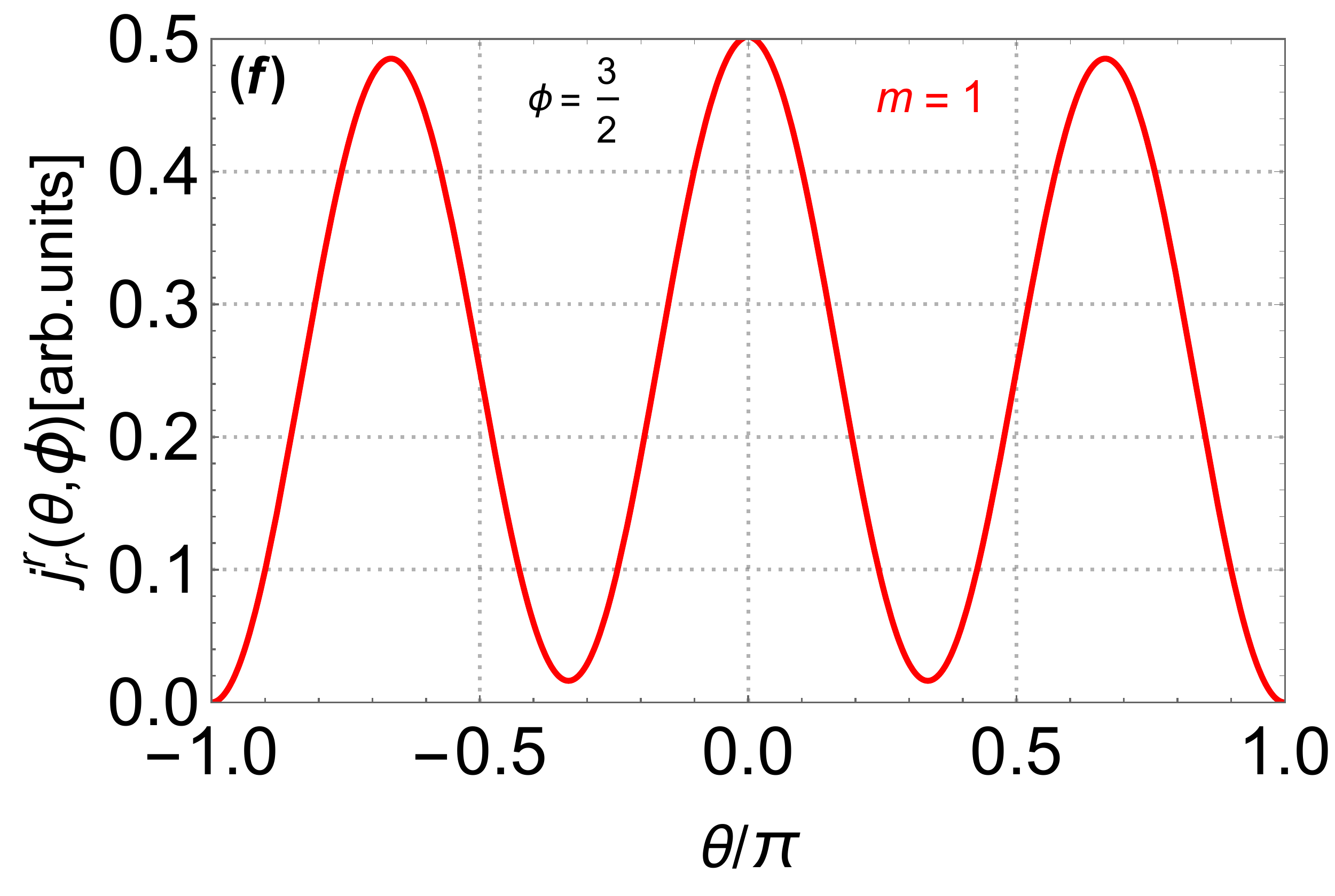}\\
\includegraphics[scale=0.25]{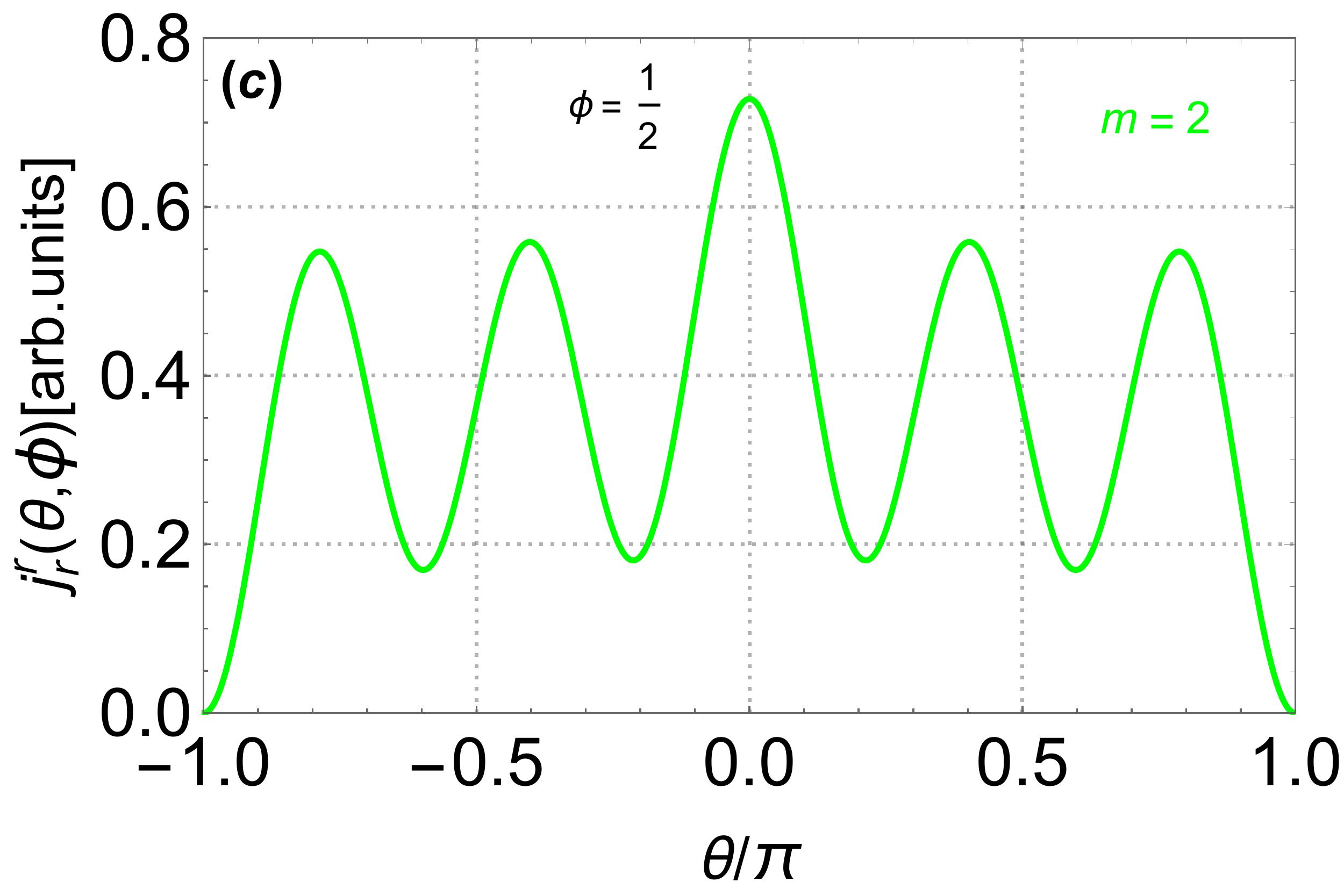}\ \ \ \ \ \ \ \ \includegraphics[scale=0.25]{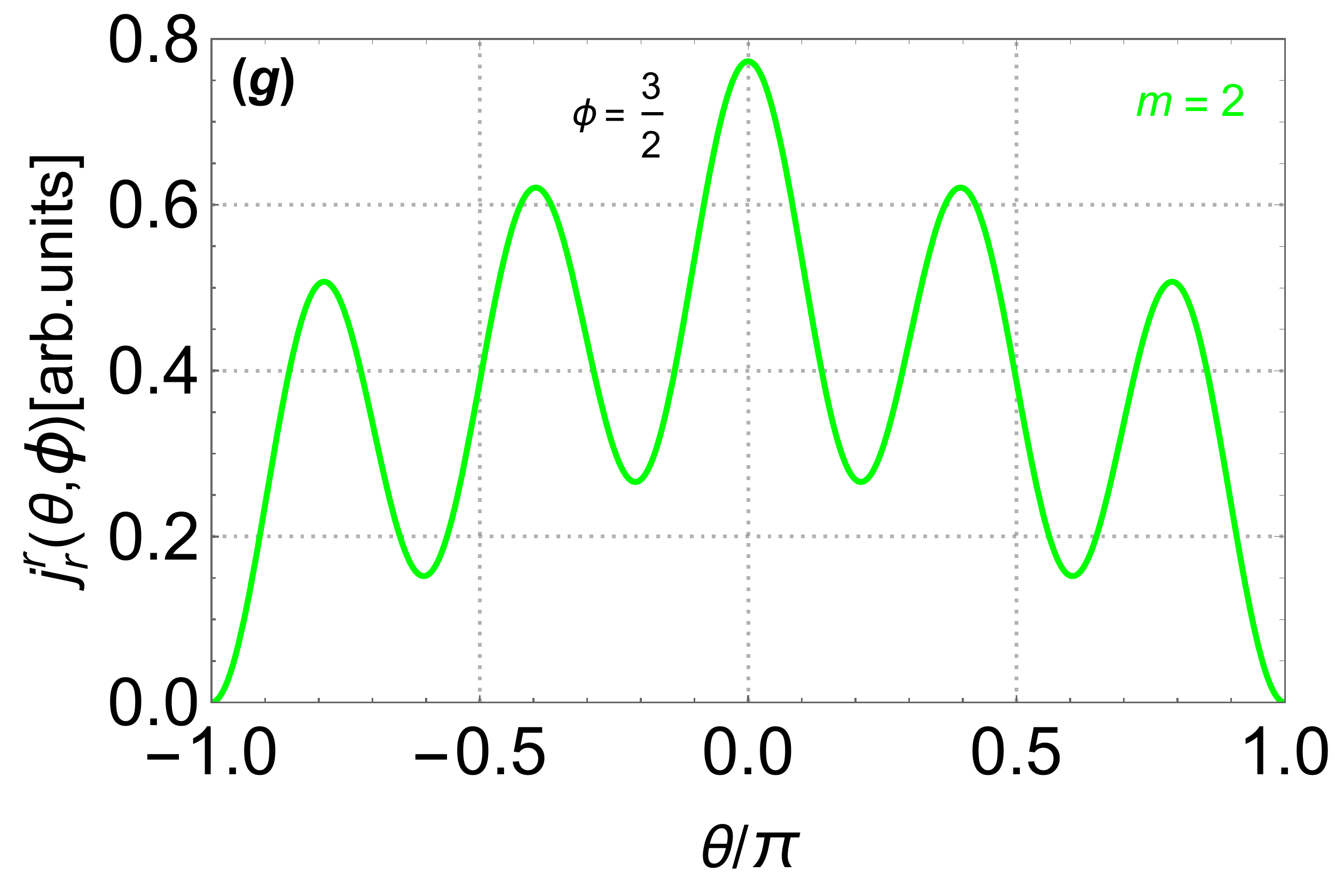} \\
\includegraphics[scale=0.25]{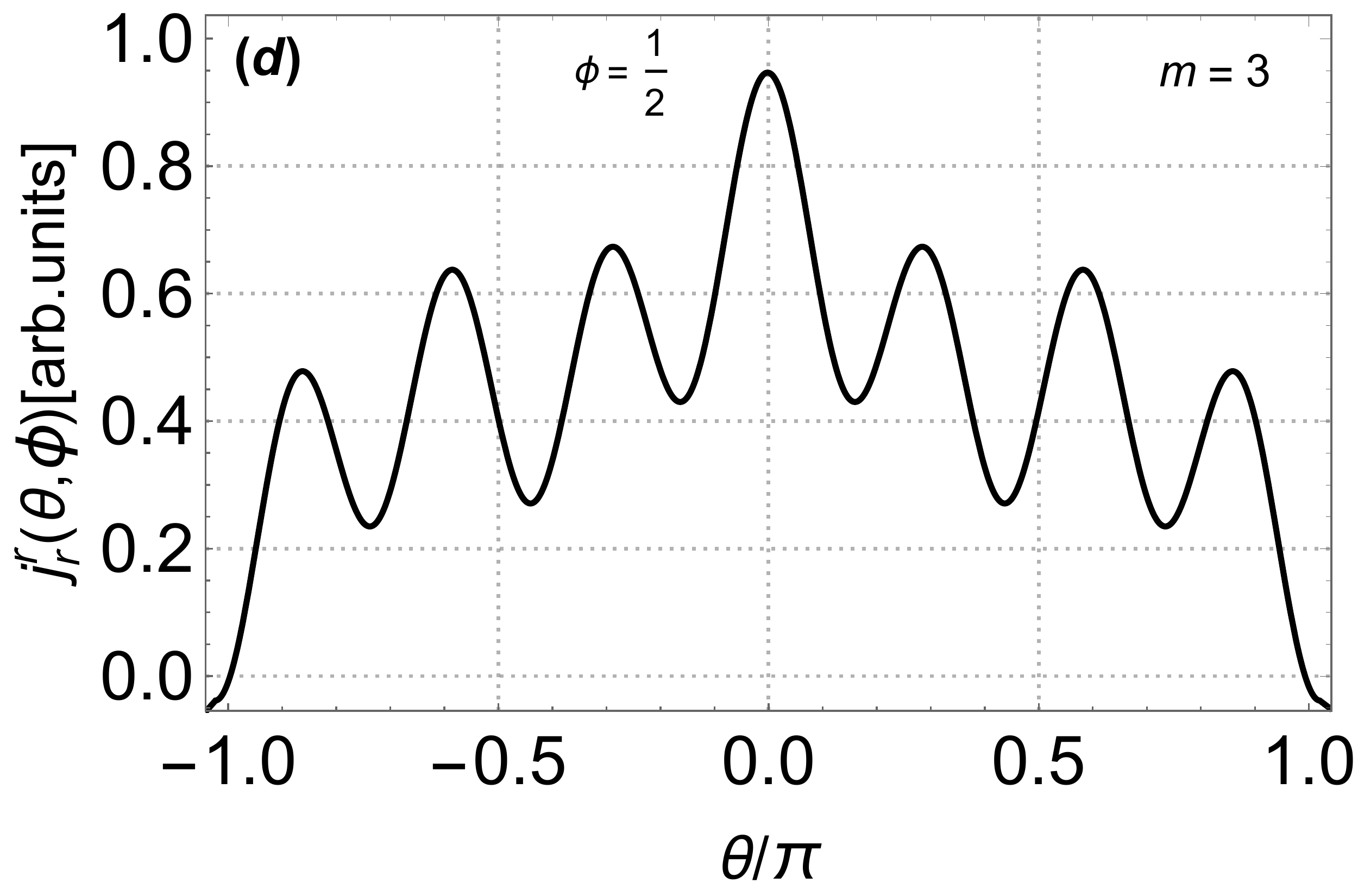}\ \ \ \ \ \ \ \  \includegraphics[scale=0.25]{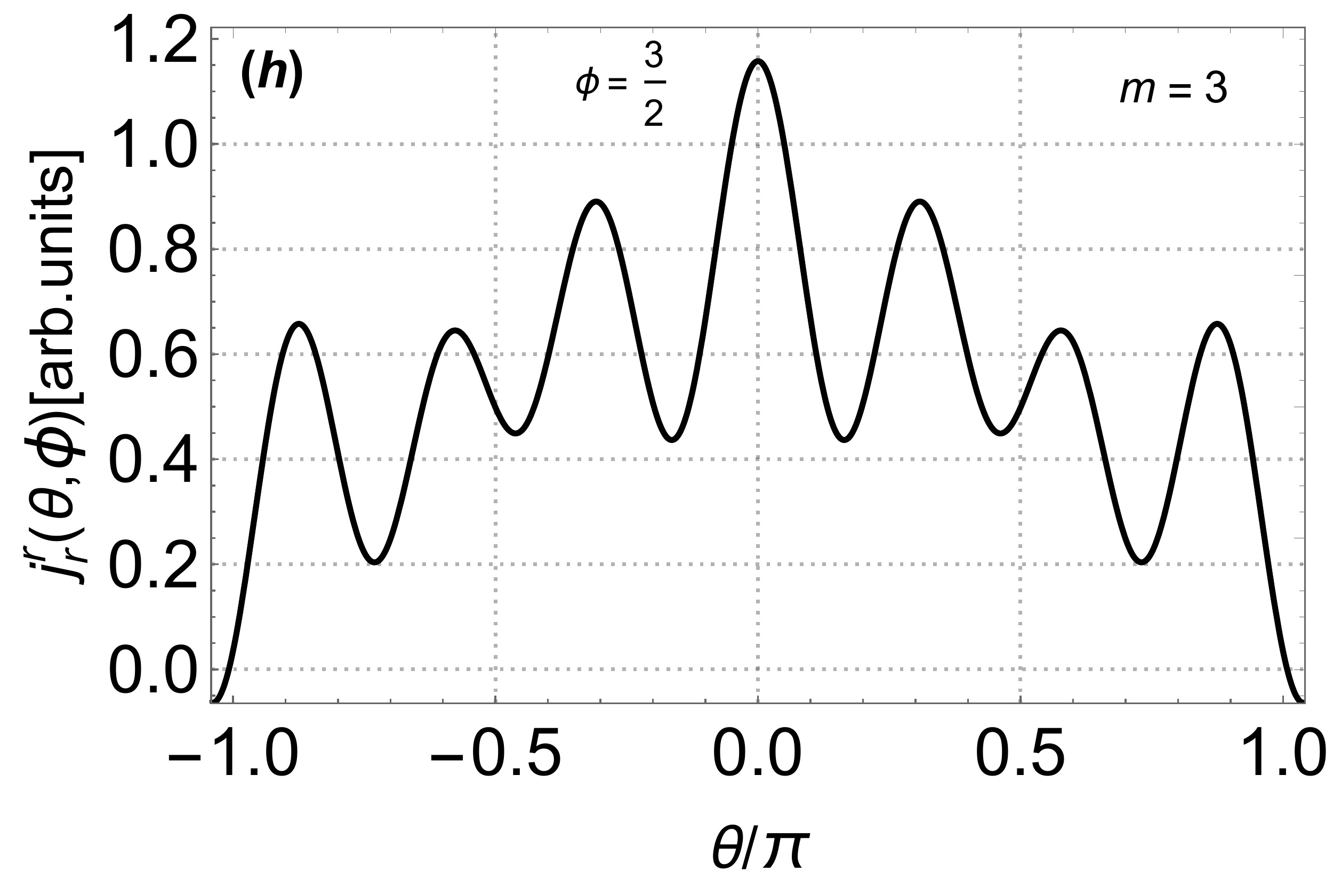}
	\caption{{\sf(color online) The radial component of the
	far-field scattered current $J^r_r$ as a function of the angle $\theta$ for  $V=1$, $\Delta=0.2$. (a,e): $E =0.0028$ and $R =3$, (b,f): $E =0.382$ and $R=5$, (c,g): $ E=0.67 $ and $ R=7.8$, (d,h): $E=0.99$ and $R=6.25$. The left panels for $\phi=\frac{1}{2}$ and  the right ones for $\phi=\frac{3}{2}$.}}\label{f8}
\end{figure}

{In Figure \ref{f8} we show the radial component of the far-field scattered current $j^r_r$ as a function of the angle $\theta$ for $\Delta=0.2$, $V=1$,  $\phi=\frac{1}{2}, \frac{3}{2}$ with some values of the incident energy $E$ and the radius $R$ of the quantum dot.
For $\theta=0$, the scattering presents a maximum and zero for
$\theta=\pm \pi$ (no backscattering). Then for each mode $m$ there is  $2m+1$ preferred directions of
 scattering observable but with different amplitudes. Otherwise, for $m=0$ (panels
 (a,e)  only forward scattering is favored, for $m=1$ panels (b,f) three
 preferred scattering directions, for $m=2$  panels (c,g) five preferred
 scattering directions and for $m=3$ panels (d,h) seven preferred scattering directions
\cite{C.Schulz2015,R.Heinisch2013}. We  observe that the mode $m=0$ is relatively large
  compared to the strong resonances of the higher modes. Furthermore, by comparing these results for
  $\phi=\frac{1}{2}, \frac{3}{2}$ with the obtained result in \cite{Belouad2018} for $\phi=0$,
we notice that the oscillatory behavior of the radial component of the far-field scattered
current $j^r_r$  for each mode $m$, not the same except for $m=0$ and their amplitude increases when $\phi$ increases.}

\begin{figure}[!hbt]
\centering
\includegraphics[scale=0.25]{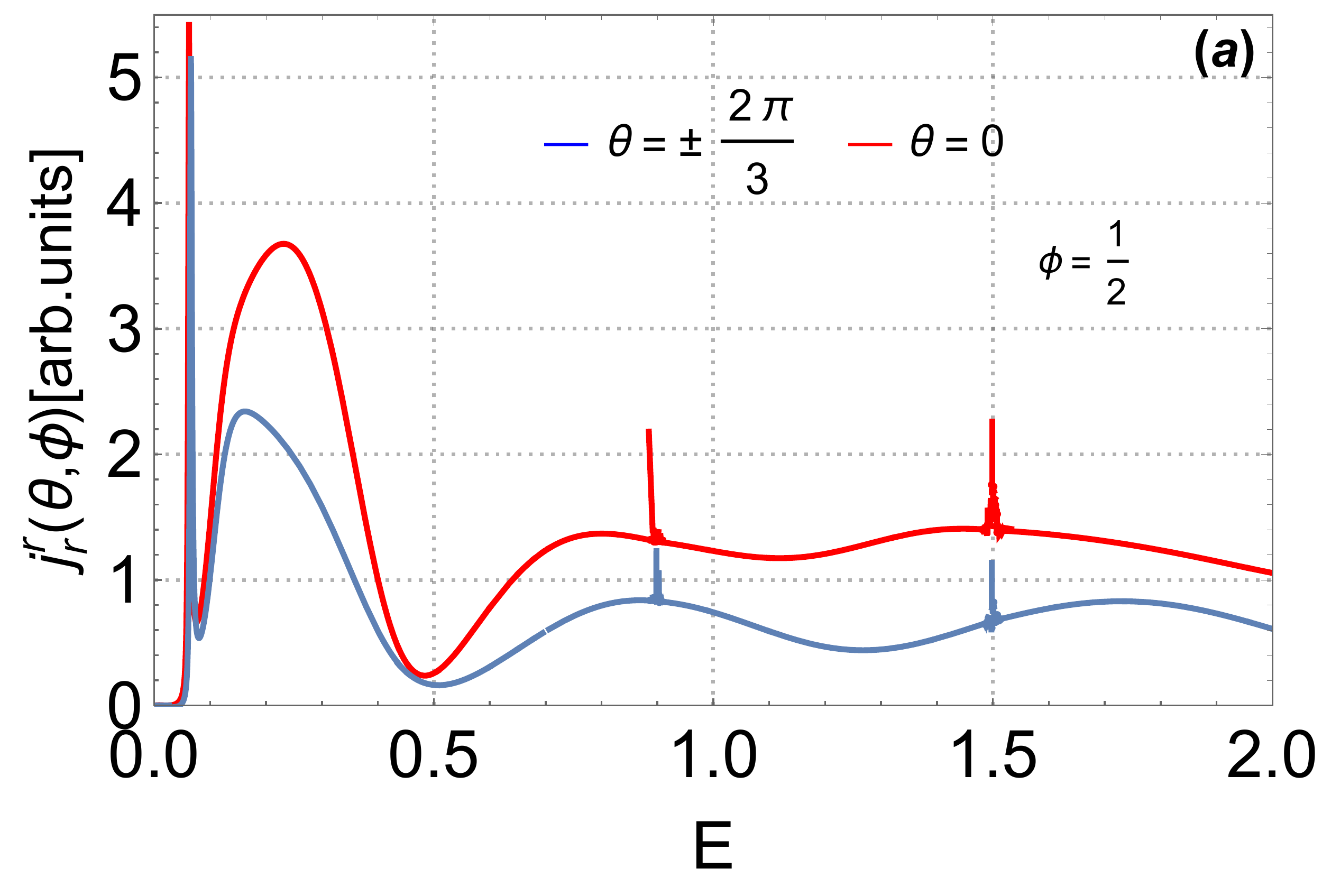}\ \ \ \ \ \ \ \  \includegraphics[scale=0.25]{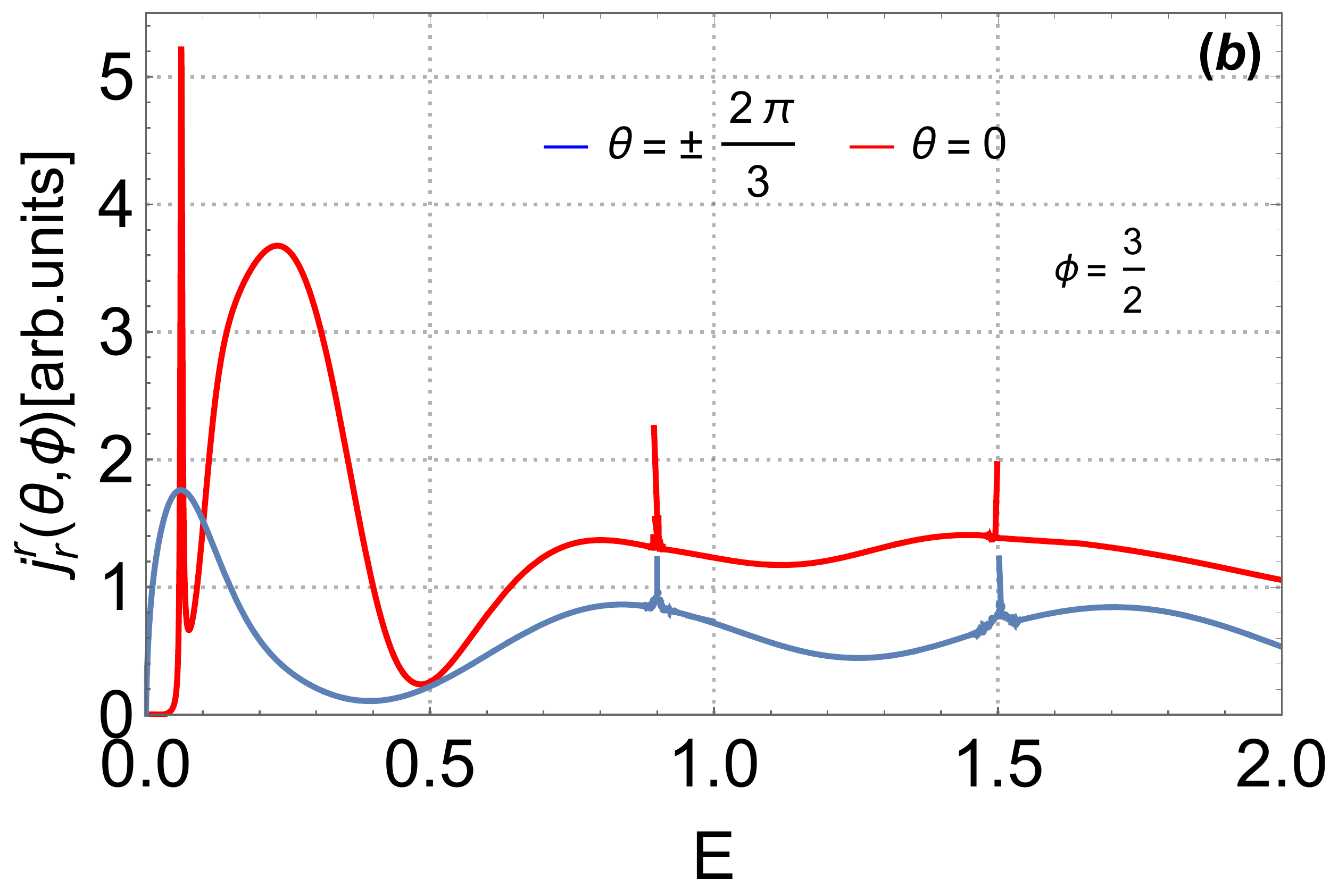}
\caption{\sf (color online) {The radial component of the far-field scattered current $j^r_r$  as a function of the incident energy $E$
for the angles $\theta=0$ (red line) and $\theta=\pm 2 \pi/3$ (blue line)
with $V=1.2$, $\Delta=0.3$,  $R=4$. (a): $\phi=1/2$  and (b): $\phi=3/2$ .
}}\label{f9}
\end{figure}

In Figure \ref{f9} we plot the radial component of the far-field scattered current $j^r_r$
as a function of the incident energy
for  three different values of the $\theta$ ($\theta=0$ and $\theta=\pm 2 \pi/3$)
 with $V=1.2$, $\Delta=0.3$, $R=4$ such that (a):  $\phi=\frac{1}{2}$  and (b):
 $\phi=\frac{3}{2}$.  Figure \ref{f9}a shows a characteristic feature of all the three values of $\theta$, which is the absence of
 $j^r_r$ for $E\rightarrow 0$. Afterward, we observe that for $E=0.05$ there is the presence of  maximum peaks of $j^r_r$ for the three values of $\phi$. While in the regime $E>0.05$,  $j^r_r$ is showing 
 oscillating behaviors, which 
 become more damped when $E$ increases.
Figure \ref{f9}b tells us that the behavior of $j^r_r$ for $\theta=0$ is similar to that found in Figure \ref{f9}a, but for $\theta=\pm 2 \pi/3$ the behavior of $j^r_r$ is quite different, equivalently to write
$j^r_r(0,\frac{1}{2})=j^r_r(0,\frac{3}{2})$ and $j^r_r(\pm \frac{2\pi}{3},\frac{1}{2})\neq j^r_r(\pm\frac{2\pi}{3},\frac{3}{2})$. At the resonance energies $E=V_-=0.9$ and $E=V_+=1.5$ we observe the appearance of different peaks.\\

\section{Conclusion}

We have studied the scattering of Dirac electrons in a quantum dot of graphene
subject to  potential barrier $V$ and energy gap $\Delta$ in the presence of the magnetic flux $\phi$. The appropriate boundary conditions for the Dirac equation have been derived and it has been illustrated how it is possible to formally use
the scattering coefficients $a_m(\phi)$ and $b_m(\phi)$ describing the characteristics of our systems. The scattering  efficiency, the square modulus of the scattering coefficient and the radial component of current density  were calculated.

The scattering of Dirac electrons was studied in three regimes of the energy of the incident electron $E<V_-$, $V_-<E<V_+$ and $V_+<E$. For the regime $E<V_-$ we have shown when the radius $R$ of the quantum dot tends towards zero, the scattering efficiency $Q$ is zero for  $\phi=\frac{1}{2}$ and presents a maximum in the form of a peak for $\phi=\frac{3}{2}$. However, when $R$ becomes important $Q$ shows a damped oscillatory behavior with the appearance of emerging peaks and the decrease of their amplitude, the small values of $E$ correspond to large amplitudes of $Q$. For the other regimes we have shown that $Q$ has a damped oscillatory behavior,
with more peak when $\phi$ increases. Subsequently, we have studied the effect of the potential $V$ on the diffusion phenomenon, we have shown that when $V$ increases $Q$ also increases. The increase in magnetic flux $\phi$ is accompanied by the appearance of peaks at the resonance points, which correspond to the energy value $E=V_\pm$. In addition, we have found that $Q$ exhibits an oscillatory behavior as a function of the energy gap $\Delta$ with the asymmetry
behavior $Q\left(\Delta,\phi=\frac{1}{2}\right)
\neq Q\left(\Delta,\phi=\frac{3}{2}\right)$.

To identify the resonances, we have analyzed the energy dependence of the square module of the diffusion coefficients $|c_m(\phi)|^2$. It was found that near $E=0$, only the lowest diffusion coefficient  is non-null but with  the increase
of $E$  the remaining coefficients started to show some  contributions. For larger $E$,  $|c_m(\phi)|^2$ tend to have oscillatory behavior but for a not too large $E$ we have seen that  the successive appearance of modes is interspersed with sudden and sharp peaks of different $|c_m(\phi)|^2$. Regarding the angular characteristic of the reflected radial component, we have found that each mode has $(2m+1)$ preferred directions of diffusion observable with different amplitudes.

\section*{Acknowledgment}
The generous support provided by the Saudi Center for Theoretical
Physics (SCTP) is highly appreciated by AJ.


\end{document}